\documentclass[twocolumn]{aastex63}
\usepackage{amsmath}
\usepackage{xcolor}
\usepackage{hyperref}
\usepackage{float}
\usepackage{siunitx}

\newcommand{\del}{$\delta\,$}
\newcommand{\Halpha}{H$\rm \alpha$ }
\renewcommand{\arraystretch}{1.25}

\begin{document}
\title{The Large Scale Behaviour in the Disk of \del Scorpii from 2000-2018}
\author{M.W. Suffak}
\affiliation{Department of Physics and Astronomy, Western University, London, ON N6A 3K7, Canada}
\author[0000-0001-9900-1000]{C.E. Jones}
\affiliation{Department of Physics and Astronomy, Western University, London, ON N6A 3K7, Canada}
\author{C. Tycner}
\affiliation{Department of Physics, Central Michigan University, Mt. Pleasant, MI 48859, USA}
\author{G.W. Henry}
\affiliation{Center of Excellence in Information Systems, Tennessee State University, Nashville, TN 37209, USA}
\author{A.C. Carciofi}
\affiliation{Instituto de Astronomia, Geof\'isica e Ci\'encias Atmosf\'ericas, Universidade de S\~ao Paulo, Brazil}
\author{B.C. Mota}
\affiliation{Instituto de Astronomia, Geof\'isica e Ci\'encias Atmosf\'ericas, Universidade de S\~ao Paulo, Brazil}
\author{A.C. Rubio}
\affiliation{Instituto de Astronomia, Geof\'isica e Ci\'encias Atmosf\'ericas, Universidade de S\~ao Paulo, Brazil}

\begin{abstract}
We model the circumstellar disk of \del Sco using the 3-dimensional Monte Carlo radiative transfer code HDUST in order to quantify the large scale changes in the disk through the years 2000 to 2018, and to see if these changes can be attributed to the secondary star affecting the disk throughout its orbit. We determine our best-fitting models through matching simulated observations to actual \Halpha spectroscopy and V-band photometric observations. Our modelling results confirm previous findings that the disk of \del Sco was forming early in the century. We also find a period of disk dissipation when the companion is at apastron, as well as a significant growth of the disk between 2009 and 2011, prior to the periastron of 2011. Due to the steady-state nature of the disk after 2011, it is difficult to say whether the variations seen are due to the effect of the close passage of the binary companion.
\end{abstract}

\section{Introduction}
\label{sec:introduction}
One of the main defining characteristics associated with Be stars is emission lines in the Balmer series. The classical definition, popularized by \cite{Collins1987}, defines a Be star as ``A non-supergiant B star whose spectrum has, or had at some time, one or more Balmer lines in emission." These lines are known to form within an outflowing gaseous circumstellar disk that has developed around the star, of which the details are not fully understood \citep{rivinius2013classical}. Another key feature of Be stars is rapid rotation, which, when coupled with pulsations, may supply the necessary energy required to act as the mass-loss mechanism of the star in order to form this disk \citep{Baade2016}, though the details remain unconfirmed. The circumstellar disks have also been characterized by the infrared continuum excess they produce \citep{Waters1986}, as well as linearly polarized light, resulting from electron scattering within the disk \citep{McDavid1990, Marr2018}.

\cite{Porter2003} described several models that have been proposed to describe these circumstellar disks. Of these models, the viscous decretion disk model, VDD, of \cite{Lee1991} has seen the most success in reproducing the observations of Be stars. While this model does not explain the mass loss mechanism that would contribute to forming the disk, it is the most widely accepted model of the circumstellar disks, and has been explained in detail in \cite{Bjorkman1997} and \cite{Bjorkman2005}, among others. The viscous decretion disk model has also been interpreted using the non-LTE Monte Carlo radiative transfer code HDUST \citep{carciofi2006non} which can provide predicted observables. This code has been used in many studies of Be stars such as \cite{Vieira2017}, \cite{Klement2015}, and \cite{Ghoreyshi2018}.

Many Be stars are also known to exist in binary systems. The survey of B and Be stars by \cite{Oudmaijer2010} found 30\% of the Be stars they observed to exist in binary systems, although some researchers have suggested that all Be stars are binaries \citep{Kriz1975}. A recent study of the radio spectral energy distribution (SED) of Be stars point to the same direction \citep{Klement2019}. \cite{Okazaki2002} studied the interaction of coplanar Be disks and neutron stars in Be/X-ray binaries through the use of a three-dimensional (3D) Smoothed Particle Hydrodynamics (SPH) code. They found a strong tidal interaction between the disk and binary companion, resulting in a phase-dependent disk structure. \cite{panoglou2016discs} and \cite{Cyr2017} also used this SPH code to simulate Be discs in binary systems for coplanar and misaligned systems. They displayed the same tidal effect found by \cite{Okazaki2002}, as well as the effect a binary star can have in highly eccentric prograde and retrograde orbits. \cite{Cyr2017} found that a \ang{30} misalignment angle of the binary orbit can cause a \ang{10} warping of the disk, while \cite{panoglou2016discs} found that highly elliptical ($e\, =\, 0.9$) prograde and retrograde orbits can cause significant density enhancements in parts of the disk, and circular binary orbits cause truncation of the disk much more strongly in the prograde case than in the retrograde case.

The binary Be star \object{\del Scorpii} (B0.5V) is known to have a companion star with a highly eccentric orbit ($e \approx 0.94$) and a period of almost 11 years \citep{tycner2011revised}. This high eccentricity brings the two binary components within 0.8 au ($\approx 25 \,  R_*$, 5.9 mas) of each other \citep{miroshnichenko20132011}, which allows the companion star to potentially affect the circumstellar disk of the primary as shown in \cite{panoglou2016discs}, who used a similar eccentricity to \del Sco in their simulations. Upper limits have been placed on the disk size during previous periastrons of $10\,  R_*$ \citep{miroshnichenko2003spectroscopy} and $20\,  R_*$ \citep{miroshnichenko20132011}, indicating the companion star may have come as close as $5\,   R_*$ to the disk in 2011. It has been shown for close circular binary stars that well-confined one-armed density modes can occur in the disks of Be stars \citep{Ogilvie2008, Oktariani2009}, and it is postulated that a similar effect may occur for eccentric binary systems  \citep{Oktariani2016}. Given the estimated spectral type of the companion as a B2V star \citep{tango2009new}, the components of the \del Sco system are of similar size and thus may produce strong interactions between themselves as well as the circumstellar disk at, or near, periastron. 

\del Sco was first classifed as a Be star when \cite{cote1993new} observed a small amount of \Halpha emission in its spectrum. Since this reclassification of \del Sco as a Be star, two periastrons have passed, once in 2000, and again in 2011. Spectroscopic observations by \cite{miroshnichenko2001spectroscopic} of \del Sco around the 2000 periastron revealed a large increase in \Halpha emission compared to that found by \cite{cote1993new}, with further noticeable month-to-month variations in its \Halpha EW and visual magnitude. They suggested these small variations were due to the disk's inability to grow greater than the Roche lobe of the primary, which caused a density increase on the side of the disk facing the secondary. There was also a large spectroscopic campaign carried out around the 2011 periastron by \cite{miroshnichenko20132011}, who used \Halpha and He II line observations to constrain the date of periastron through radial velocity measurements. They also utilized the evolutionary models of \cite{Ekstrom2012} to confirm their masses for the binary components as well as the age of the \del Sco system. The photometric behaviour around the previous periastrons has also been monitored by \cite{otero2001optical}, who noticed a significant brightening in the V-band around the 2001 periastron, and by \cite{jones2013using}, who utilized Johnson BV photometry over the period of 2009 to 2012 to find significant long-term disk building events, as well as shorter cyclical variability on the order of 60 to 100 days.

There have been numerous other efforts in modelling the disk of \del Sco. \cite{carciofi2006properties} utilized continuum flux and polarization modelling to find the best-fit disk parameters for their 2001 spectropolarimetry data. \cite{millan2010spectro}, \cite{meilland2011binary} and \cite{che2012imaging} have all utilized interferometric measurements to model the disk of \del Sco and monitor the extent of its emitting region through fitting models to their observations of the spectral lines Br$\gamma$, He I, and H$\alpha$ in conjunction with measurements in the H and K band continuum. Despite the inability of most models to model non-coherent scattering, which contributes significantly to the wings of emission lines, the modelling of spectral lines, particularly the \Halpha line, has been used in the study of many Be star systems since the \Halpha line is the most prominent in the spectrum of a Be star. For example, \cite{Silaj2016}, \cite{Jones2008} and \cite{Jones2017} modelled \Halpha lines to determine the disk properties of the Be stars  48 Lib, $\kappa$ Dra, $\beta$ Psc, $\nu$ Cyg, and 48 Per.

It is the aim of this paper to model the large scale, long-term changes of the disk of \del Sco by constraining our models with \Halpha spectroscopy and V-band photometric measurements. We look to determine whether these changes may be due to the binary companion of \del Sco interacting with the disk of the primary. Our methodology is presented in Section \ref{sec:methodology}, observations and collected data is shown in Section \ref{sec:observations}, and our results and discussion are, respectively, presented in Sections \ref{sec:results} and \ref{sec:discussion}.

\section{Methodology}
\label{sec:methodology}

\subsection{BeAtlas}
\label{sec:BEMCEE}
BeAtlas, developed by \cite{Mota2019}, consists of two systematic grids of Be star models computed by the code HDUST (see section \ref{sec:HDUST}). The first is a photospheric (diskless) grid, and the second is a grid involving the star and the disk. Here we only work with the photospheric grid, which contains 7700 models of spectral type O8 to A7. The grid varies the stellar mass from 1.7 to 20 $\rm M_\odot$, rotation rate from 0 to 0.99 times the critical velocity, stellar age from 0 to 1.02 times the length of the main sequence of the star, and inclination from \ang{0} to \ang{90}.

\cite{foreman2013emcee} developed an algorithm for a Markov chain Monte Carlo (MCMC) sampling method for determining the posterior probabilities of a set of model parameters given a set of data. This algorithm has been adapted and implemented by \cite{Mota2019} as a statistical tool to explore the BeAtlas grid, for the purposes of finding the probability distribution function (PDF) of the stellar parameters given the observed data. The MCMC algorithm employs a defined number of random processes, or ``walkers," taking a defined number of random ``steps" to explore the parameter space of the mass, rotation rate, age, inclination, distance, and reddening of these models. Since the number of models in the BeAtlas grid is finite, the models are interpolated between in order to allow full coverage of the parameter space. At each step, the code computes the posterior probability function defined by the sum of the likelihood function and the prior function. The likelihood function is defined as
\begin{equation}
   \log p(D|\Theta,\alpha) = -0.5\bigg[\frac{\log(F_{obs}/F_{mod})}{\sigma_{F_{obs}}/F_{obs}}\bigg]^2,
\label{eq:likelihood_eqn}
\end{equation}
where $D$ represents the observational data, $\Theta$ is the model parameters, $F_{obs}$ and $F_{mod}$ are the observed and model fluxes respectively, $\sigma_{F_{obs}}$ is the error of the observed fluxes, and $\alpha$ is the set of nuisance parameters, which is required to model the process that generates the data, but is otherwise of no interest \citep{foreman2013emcee}. As well, the prior function may be used when some information about a given parameter is known beforehand. For instance, with the distance parameter, an observed parallax may be used as a prior through the formula
\begin{equation}
   \log p_{\pi}(\Theta, \alpha) = -0.5 \bigg(\frac{\pi-\pi_{mod}}{\sigma_\pi}\bigg)^2,
\label{eq:prior_eqn}
\end{equation}
where $\pi_{mod}$ is the model parallax, and $\pi$ and $\sigma_\pi$ are the observed parallax and error, respectively. Clearly, in Equation \ref{eq:prior_eqn} it is assumed that the parallax follows a Gaussian distribution. This prior function is equally applicable to the other explored parameters. The program that employs the procedure described above is known as BEMCEE \citep{Mota2019}.

BeAtlas was built using the Geneva stellar evolution models of \cite{Georgy2013} that allow the conversion of the fundamental stellar parameters (mass, rotation rate, age) into derived parameters such as polar radius and luminosity (see \citealt{Mota2019} for more details). This process will find the stellar parameters for the primary star of \del Sco that we can compare with other values in the literature.

\subsection{HDUST}
\label{sec:HDUST}

The code HDUST \citep{carciofi2006non} is a non-local thermodynamic equilibrium (NLTE) Monte Carlo radiative transfer code capable of predicting observables from 3D circumstellar disk models. HDUST uses Monte Carlo routines to find the hydrogen ionization fraction and level populations, as well as determine a self-consistent temperature structure for the disk. This information is used to produce simulated observations, such as the SED, spectral lines, and polarization of the star/disk system over desired wavelengths. 

In this work, we use HDUST to model the disk of \del Sco using a power law for volume density $\rho(r,z)$ within Be star disks
\begin{equation}
    \rho(r,z) = \rho_0 \bigg(\frac{R_*}{r}\bigg)^n \exp\bigg(-\frac{z^2}{2H^2}\bigg),
\label{eq:volume_density}
\end{equation}
with $r$ and $z$ respectively being the radial and vertical positions within the disk, $R_*$ as the equatorial radius of the star, $\rho_0$ as the density where $r\, = \,R_*$ and $z\, =\, 0$, $n$ being a parameter which defines how quickly the density decreases with increasing $r$, and the scale height $H$ is defined as
\begin{equation}
    H(r) = H_0\bigg(\frac{r}{R_*}\bigg)^\beta,
\label{eq:scale_height}
\end{equation}
where
\begin{equation}
    H_0 = \frac{a}{v_{crit}}R_*,
\label{eq:H_not}
\end{equation}
with $a$ being the sound speed of the disk, which is determined within the code from the temperature structure, and $v_{crit}$ is the critical rotational velocity of the star. The disk flaring exponent $\beta$ is set to 1.5, which is the number used for roughly isothermal, optically thin disks \citep{carciofi2006non}. However, \cite{carciofi2006non} also show that varying this $\beta$ exponent has no visible effect on the temperature structure of the disk, due to the inner disk region being insensitive to $\beta$ and the outer portion being optically thin, so we choose to keep $\beta\, =\, 1.5$ constant for all models.

We vary the parameters $\rho_0$ and $n$ to produce our models. It is important to note here that $n$ is thought to embody the state of the disk. \cite{Vieira2017} shows that, for $n\,\lesssim \,3$, the disk is dissipating, for $3\,\lesssim \,n\,\lesssim \,3.5$ the disk is in a steady state, and for $n\,\gtrsim \,3.5$ the disk is thought to be building. However there is evidence that this is not necessarily true for late-type Be stars, as Rubio (2020, in prep.) and Granada et al. (2020, in prep.) have found $n\, <\, 3$ for their steady-state late-type Be stars.

\section{Observations and Data}
\label{sec:observations}

\subsection{Spectroscopy}
\label{sec:spectroscopy}

The Ritter Observatory in Toledo, Ohio, USA, observed \del Sco using a 1 m telescope coupled with an \'echelle spectrograph and a Wright Instruments Ltd. CCD camera during the 2000 to 2003 time period. These observations have a resolution of $ R \, = \, 26,000$ and cover the wavelength region $\rm 5285\text{ to }6600\, $\AA. We retrieved some of these spectra from the Ritter Observatory Public Archive, as well as the EW data from \cite{miroshnichenko2003spectroscopy} where the observations were originally published.

\Halpha spectra were also obtained from the Canada-France-Hawaii Telescope (CFHT) Science Archive via the Canadian Astronomy Data Centre (CADC). The CFHT (in Maunakea, Hawaii) collected \Halpha spectra of \del Sco in the years 2007, 2010, 2011, 2016, and 2017, through use of their Echelle Spectropolarimetric Device for the Observation of Stars at CFHT (ESPaDOnS). This instrument has a resolving power of $\rm R\, = \, 65,000$.

Additional \Halpha spectra were obtained using the fiber-fed \'echelle spectrograph attached to the 1.1 meter John S. Hall telescope at the Lowell Observatory in Flagstaff, Arizona. We have 101 spectra from 2005 to 2018 that were taken with this instrument at a resolving power of $R\, = \, 10,000$. These observations have been made available online in a machine readable table. Table \ref{tab:Ha_spectra} gives the first three rows of this table for guidance.

\begin{deluxetable}{cccc}[ht!]
\tablecaption{\Halpha observations for \del Sco from the Lowell Observatory. The table indicates (left to right): the spectrum number, the modified Julian date of the observation, wavelength ($\lambda$, in nm), and the ratio of normalized flux over continuum flux (F/$\rm F_c$).\label{tab:Ha_spectra}}
\startdata
\\
Spectrum & MJD (+2400000.5) & $\lambda$ & F/$\rm F_c$\\[1pt]
\tableline
1 & 53461.442 & 648.183655 & 0.991100 \\[1pt]
1 & 53461.442 & 648.207947 & 0.981596 \\[1pt]
1 & 53461.442 & 648.232300 & 0.970491 \\[1pt]
\enddata
\tablecomments{Table \ref{tab:Ha_spectra} is published in its entirety in the machine-readable format. A portion is shown here for guidance regarding its form and content.}
\end{deluxetable}

\begin{figure*}[ht!]
    \plotone{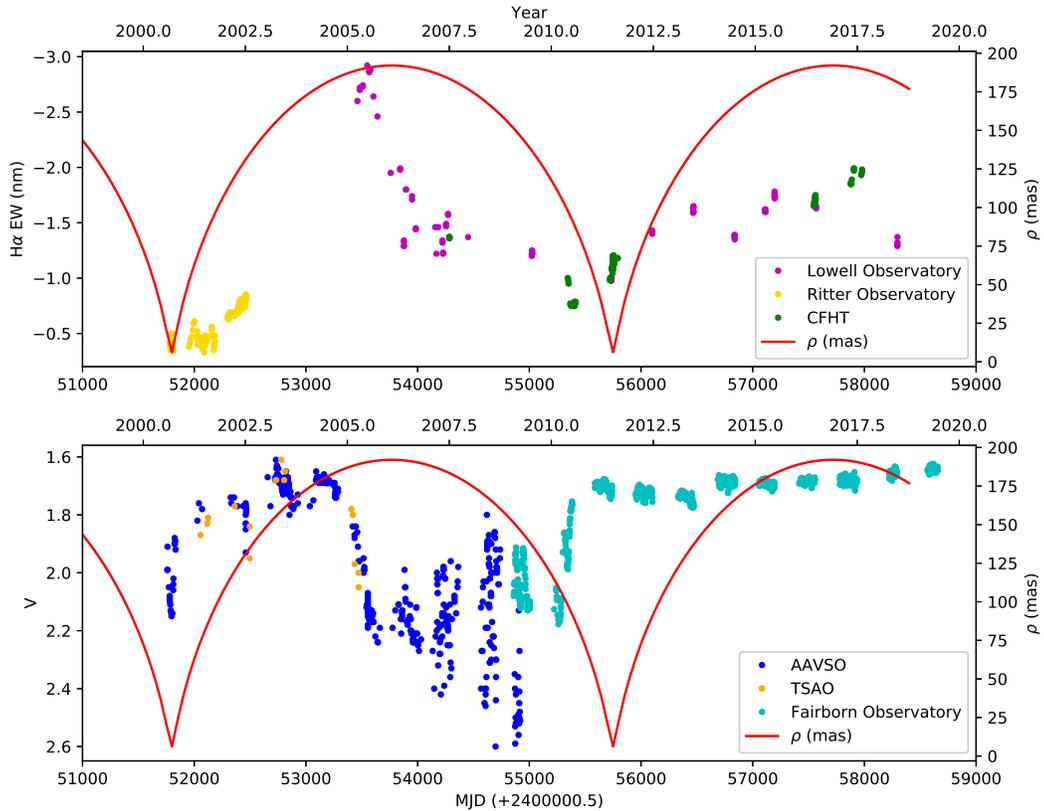}
    \caption{Plot of all H$\alpha$ EW (top plot) and V-band (bottom plot) data collected for \del Sco, along with the separation (red) between the secondary and the primary star. In the top and bottom plots, the left axis is the scale for the \Halpha EW and V magnitude, respectively, while the right axis in both plots is the scale for the separation of the two binary components in mas. The bottom x-axis is the time scale in Modified Julian Date (MJD), and the top x-axis is the time scale in years. The legend in both plots indicates the source of the EW and V magnitude data.}
    \label{fig:V_EW_plot}
\end{figure*}

The top panel of Figure \ref{fig:V_EW_plot} shows the \Halpha EW data collected for \del Sco spectra. The red line indicates the distance between the primary and secondary stars in units of mas. These distances were calculated using a program created in Mathematica\textsuperscript{TM} that utilized the orbital parameters from \cite{tycner2011revised}. The ordinate for the \Halpha EW and binary separation is on the left and right side, respectively. The legend indicates which data come from which source discussed above. The entire table of \Halpha EW data for \del Sco is available online in machine readable form. Table \ref{tab:Ha_data} gives the first three rows of this table for guidance.

\begin{deluxetable}{ccc}[ht!]
\tablecaption{\Halpha EW data for \del Sco. The table indicates (left to right): the modified Julian date of the observation, the \Halpha EW of the observation, and the telescope which took the observation.\label{tab:Ha_data}}
\startdata
\\
MJD (+2400000.5) & \Halpha EW & Source\\[1pt]
\tableline
51777.05 & -0.41 & Ritter \\[1pt]
51781.08 & -0.39 & Ritter \\[1pt]
51788.05 & -0.35 & Ritter \\[1pt]
\enddata
\tablecomments{Table \ref{tab:Ha_data} is published in its entirety in the machine-readable format. A portion is shown here for guidance regarding its form and content.}
\end{deluxetable}

\begin{figure}[ht!]
    \centering
    \includegraphics[scale = 0.3]{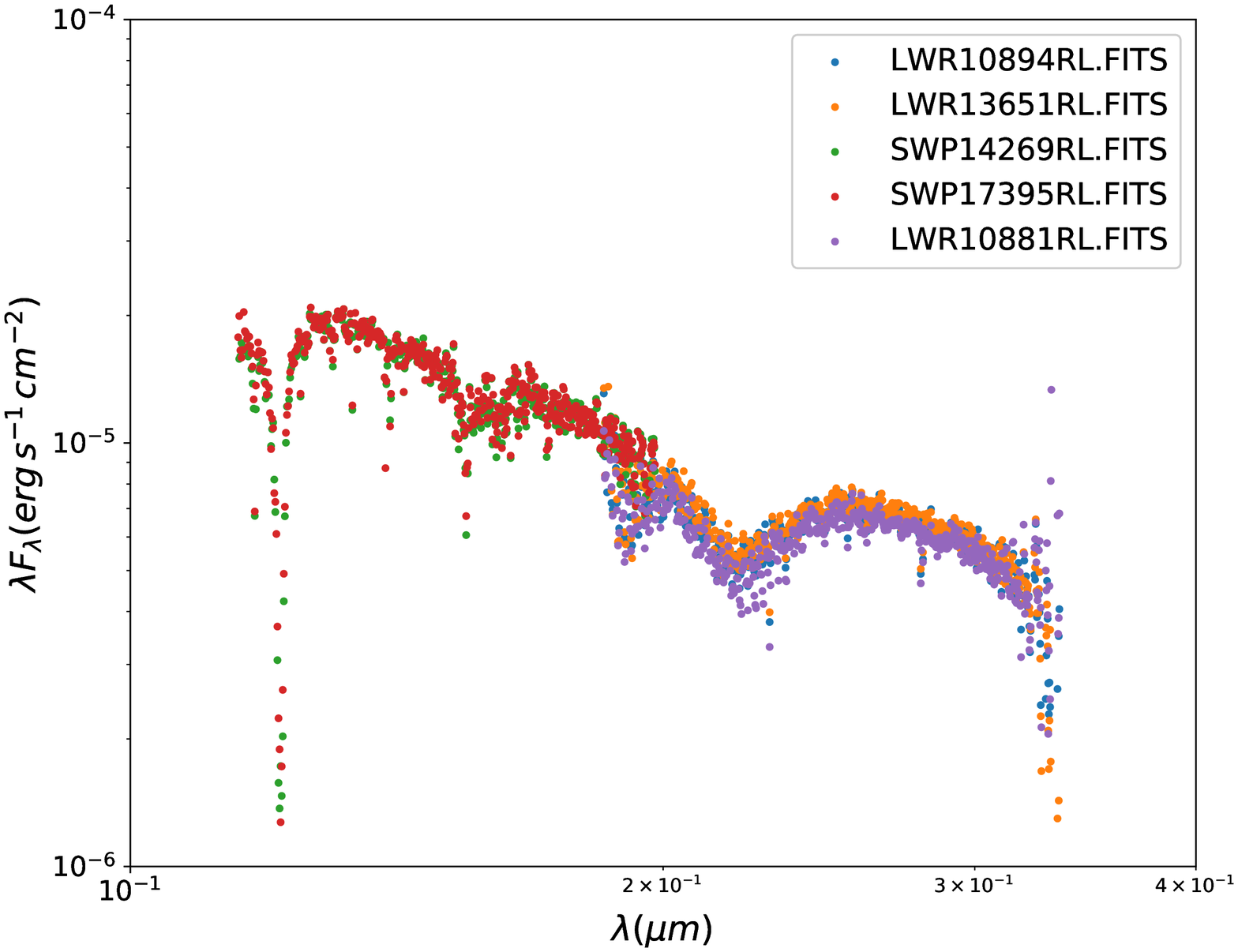}
    \caption{UV spectra for \del Sco obtained by the IUE telescope during the years 1981 and 1982. The legend shows the file names from the IUE archive website.}
    \label{fig:IUE_FULL}
\end{figure}

In addition to our \Halpha spectra, the International Ultraviolet Explorer (IUE) obtained ultraviolet (UV) spectra of \del Sco before it showed presence of a disk. The IUE was equipped with two apertures; the large aperture being approximately 10 by 20 arcseconds in diameter and the small aperture being 3 arcseconds in diameter. These apertures were used with a long-wavelength spectrograph ($1850\text{ to }3300\,$\AA) and a short-wavelength spectrograph ($1150\text{ to }2000\, $\AA). The image quality of the IUE telescope results in about a 3 arcsecond image, thus observations that used the small aperture may have some light loss, making the large aperture much more reliable \citep{IUE}. Therefore, for \del Sco we selected only the spectra using the large aperture and high dispersion settings which have a resolution of approximately 0.2 \AA $\,$ \citep{IUEesa}. We obtained 3 long-wavelength spectra, two from 1981 and one from 1982, and 2 short-wavelength spectra, one each from 1981 and 1982. These UV spectra, shown in Figure \ref{fig:IUE_FULL}, are valuable input into BEMCEE for determination of the stellar parameters.

\subsection{Photometry}
\label{sec:photometry}

Photometry for \del Sco was collected and used by \cite{carciofi2006properties}. These observations were taken in 2002 to 2005 at the Tien-Shan Observatory (TSAO) in Kazakhstan with a 50 cm telescope and a standard pulse-counting single-channel photometer. Here we will use their V-band data.

The American Association of Variable Star Observers (AAVSO) also has many photometric measurements. We collected the V-band measurements for \del Sco in the AAVSO International Database, which spans the years 2000 to 2009. We selected only the data that had the ``verified" flag attached to it, to ensure the photometry was of good quality.

Over a thousand V-band observations of \del Sco were acquired using the T3 0.4 m automatic photoelectric telescope (APT) at Fairborn Observatory in southern Arizona over the period of 2009 to 2019. T3 is equipped with a precision photometer that uses an EMI 9924B photomultiplier tube for succesive measurements of photon count rates through Johnson B and V filters. The precision of a single observation on a good night is approximately $0.003\text{ to }0.005$ mag. See \cite{Henry1999} for more details on the operation of the APT and reduction of the data.

The bottom panel of Figure \ref{fig:V_EW_plot} shows the V magnitude data (blue) we have collected for \del Sco from 2000 to 2019. It is plotted with the separation distance between the primary and secondary stars (red) as in the top portion of the Figure. The scale for V magnitude and separation is on the left and right ordinate, respectively. The data from the three photometry sources discussed above are coloured according to the legend. The entire table of V-band data for \del Sco is available online in machine readable form. Table \ref{tab:V_data} gives the first three rows of this table for guidance.

\begin{deluxetable}{ccc}[htb!]
\tablecaption{V-band data for \del Sco. The Table indicates (left to right): the modified Julian date of the observation, the V magnitude of the observation, and the source of the data.\label{tab:V_data}}
\startdata
\\
MJD (+2400000.5) & V & Source\\[1pt]
\tableline
51758.85 & 1.99 & AAVSO \\[1pt]
51760.60 & 1.91 & AAVSO \\[1pt]
51761.73 & 1.99 & AAVSO \\[1pt]
\enddata
\tablecomments{Table \ref{tab:V_data} is published in its entirety in the machine-readable format. A portion is shown here for guidance regarding its form and content.}
\end{deluxetable}

\begin{figure}[ht!]
    \plotone{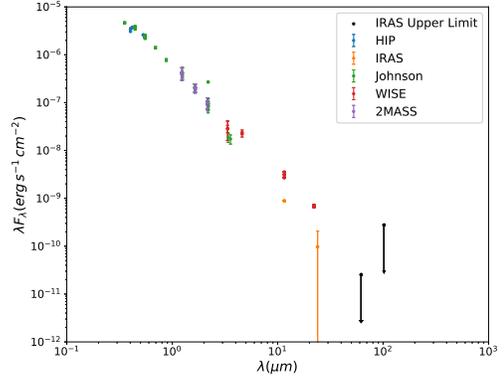}
    \caption{SED for \del Sco obtained from Vizier with a 1 arcsecond radius around the position of \del Sco. The IRAS points at 60 and 100 $\mu m$ are upper limits as indicated by the downward arrows, but are kept on the plot due to lack of other points at these wavelengths. The points are coloured according to the photometric system that was used for the observation.}
    \label{fig:CDS_plot}
\end{figure}

We also obtained an SED from the Vizier \footnote{\url{http://vizier.u-strasbg.fr/vizier/sed/}} website, which contains photometric measurements in a variety of passbands from many different sources, which have their observation coordinates within 1 arcsecond of \del Sco. The plot of these points is shown in Figure \ref{fig:CDS_plot}, where we have coloured the points according to the origin of the photometric data. Some of these data points did not have errors listed on the Vizier website, so for those points we calculated the standard deviation of the mean for each specific wavelength band and used it as an error estimate. It should also be noted that the two IRAS measurements at 60$\rm \, \mu m$ and 100$\rm \, \mu m$ are upper limits, as indicated by the downward arrows.

\section{Results}
\label{sec:results}

\subsection{Stellar Parameters}
\label{sec:stellar_params}

To find the stellar parameters for the primary of \del Sco, we take advantage of the large amount of data prior to 2000, when the star did not have a disk. We used the program BEMCEE (section \ref{sec:BEMCEE}) to find the stellar parameters for the primary of \del Sco. Since BEMCEE uses a photospheric grid of models from BeAtlas to find the stellar parameters, the SED data obtained from Vizier was searched to find data points that were taken prior to 2000. This results in data obtained with Johnson, IRAS, and Hipparcos filters available to model the SED in the diskless phase. The IRAS data, however, is outside the wavelength range of the computational grid of BeAtlas we used here. The IRAS data would not change our results greatly since at these wavelengths there is very little flux from the diskless star, thus it is not included in the program. As well, the secondary star needs to be accounted for in this fitting. The spectral type of the secondary as a B2V means the two stars are of similar brightness, and it is unrealistic to believe the photometry measurements and UV spectra contain only the light from the primary star. Thus, we subtract 20\% of the flux values from our diskless SED data as well as the IUE UV spectra to remove the effect the secondary star has on these measurements. We note that this is consistent with \cite{carciofi2006properties}, who subtracted 15\% of the dereddened flux from their observations to construct their diskless SED. As well, due to the uncertainty of the secondary being of B2 spectral type, we point out that subtracting 10\% of the flux gives nearly the same results as subtracting 20\%. So we cannot confirm the spectral type of the companion as B2 exactly.

The data set we used as input for BEMCEE consists of the IUE UV data and diskless SED points after subtracting 20\% of the flux for the secondary star. For the UV spectra, we also cut the data at an upper limit of 3000 \AA $\,$ due to noise and only selected points of good quality as was indicated in the FITS files which contain the data. Additionally, we use priors of $ v\sin i\, =\, 148\,\pm\,8\, \rm km\, s^{-1}$ \citep{Brown1997}, parallax of  $\rm 7.4\,\pm\,0.2\,mas$ \citep{tycner2011revised}, and inclination of $\rm \ang{38}\,\pm\,\ang{5}\,$ \citep{carciofi2006properties}.

\begin{figure*}[htb!]
\boxedfig{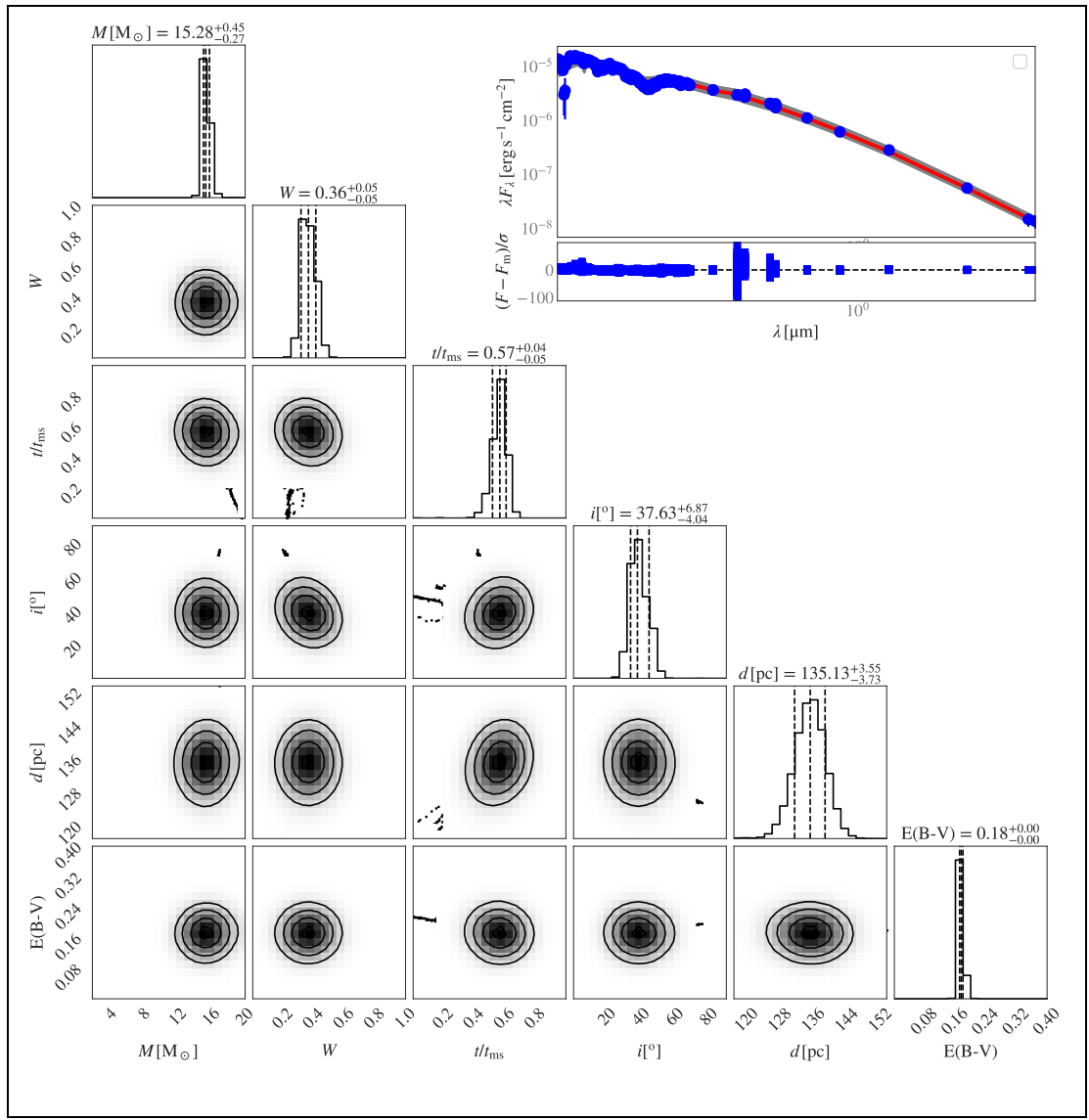}{0.8\textwidth}{}
    \caption{BEMCEE results using the data discussed previously as well as priors of $v\sin i$, parallax, and inclination. The parameters (left to right across the bottom) are mass, critical fraction (as defined in \citealt{rivinius2013classical}), time elapsed over total time of the main sequence lifetime, inclination, distance, and reddening. The top right portion of the figure shows a set of sampled models in the last step of the simulation (orange) plotted with the input data points (blue). Directly below this shows the residuals between the models and the input points. The corner plot shows the PDF for each of the 6 parameters.}
    \label{fig:Best_BEMCEE_result}
\end{figure*}

Figure \ref{fig:Best_BEMCEE_result} shows our BEMCEE result with this selected data and these priors. The six parameters explored by BEMCEE are shown in the corner plot which shows their associated PDFs. The parameters are (left to right across the bottom): mass, $M$, critical rotation fraction (as defined in \citealt{rivinius2013classical}), $W$, time elapsed over total time of the main sequence lifetime, $t/t_{ms}$, inclination, $i$, distance, $d$, and reddening, E(B-V). We defined the most likely values of each parameter as the median of the PDFs, and the uncertainties as the 16th and 84th quantiles. These are shown on the top of each column and also as the dashed lines in the diagonal plots. The top right portion of Figure \ref{fig:Best_BEMCEE_result} shows a set of sampled models in the last step of the simulation (orange) plotted with our input data points (blue). The panel below this plot shows the residuals calculated between the models and the input data points.

\begin{deluxetable}{cc}[htb!]
\tablecaption{Stellar parameters for \del Sco determined from Figure \ref{fig:Best_BEMCEE_result}.\label{tab:Best_BEMCEE_params}}
\startdata
\\
$M$ & $15.3 \, \pm \, 0.2\,\rm M_\odot$ \\[1pt]
$W$ & $0.36 \, \pm \, 0.03\,$ \\[1pt]
$ t/t_{ms}$ & $0.57 \, \pm \, 0.03\,$ \\[1pt]
$i$ & $\ang{38}\substack{+\ang{7} \\ -\ang{4}}$ \\[1pt]
$d$ & $135 \, \pm \, 4\,$ pc \\[1pt]
$E(B-V)$ & $0.18 \, \pm \, 0.01\,$ \\[1pt]
$ R_p$ & $6.4 \, \pm \, 0.2\, \rm R_\odot$ \\[1pt]
$ R_{eq}$ & $6.8 \, \pm \, 0.2\, \rm R_\odot$ \\[1pt]
$L$ & $28000 \, \pm \, 1000\, \rm L_\odot$ \\[1pt]
$ T_{eff}$ & $29700 \, \pm \, 200\,$ K \\[1pt]
$age$ & $7.3 \, \pm \, 0.4\,$ Myr \\[1pt]
$ \log (g)$ & $4.01 \, \pm \, 0.02\, \, \rm log(cm\, s^{-2})$ \\[1pt]
\enddata
\end{deluxetable}

By interpolating between the stellar evolutionary tracks of \cite{Georgy2013}, we can use the estimated values of mass, $W$, and $t/t_{ms}$ from Figure \ref{fig:Best_BEMCEE_result} to retrieve estimates of the polar radius, $R_p$, luminosity, $L$, and age of \del Sco. We can then use these parameters to find other parameters such as the effective temperature, $T_{\rm eff}$, $\log (g)$, and equatorial radius, $R_{\rm eq}$. We find the errors on these parameters by using all possible combinations of $M$, $W$, and $t/t_{ms}$ to a precision of 0.01. This yields 8030 combinations of stellar parameters for the primary star of \del Sco. From these we calculate the deviation from our estimates of the model parameters, which is used then as our error. We have not adjusted the errors for inclination, distance, or E(B-V) as they were not included in computing the other parameters from the Ekstrom models. The full list of determined parameters for \del Sco is shown in Table \ref{tab:Best_BEMCEE_params}. These parameters agree with past studies of \del Sco. Our mass is in agreement with \cite{tango2009new}, the inclination agrees with \cite{carciofi2006properties} and \cite{miroshnichenko20132011}, the distance also agrees with that of \cite{tycner2011revised} and \cite{miroshnichenko20132011}, and the value of E(B-V) is very close to the value of 0.17 in \cite{Welty&Hobbs1991}.

\begin{figure*}[htb!]
    \gridline{\fig{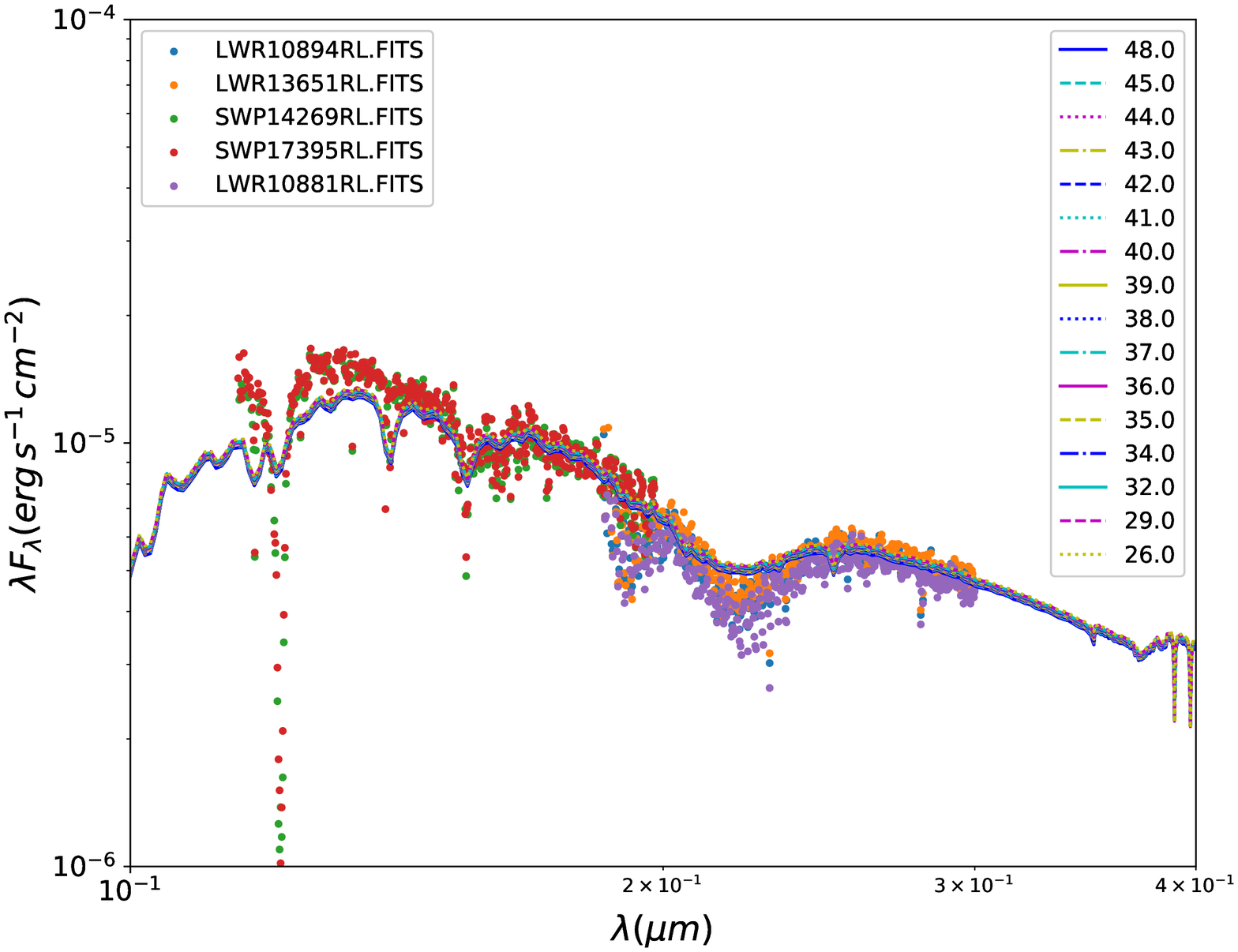}{0.4\textwidth}{(a)}
              \fig{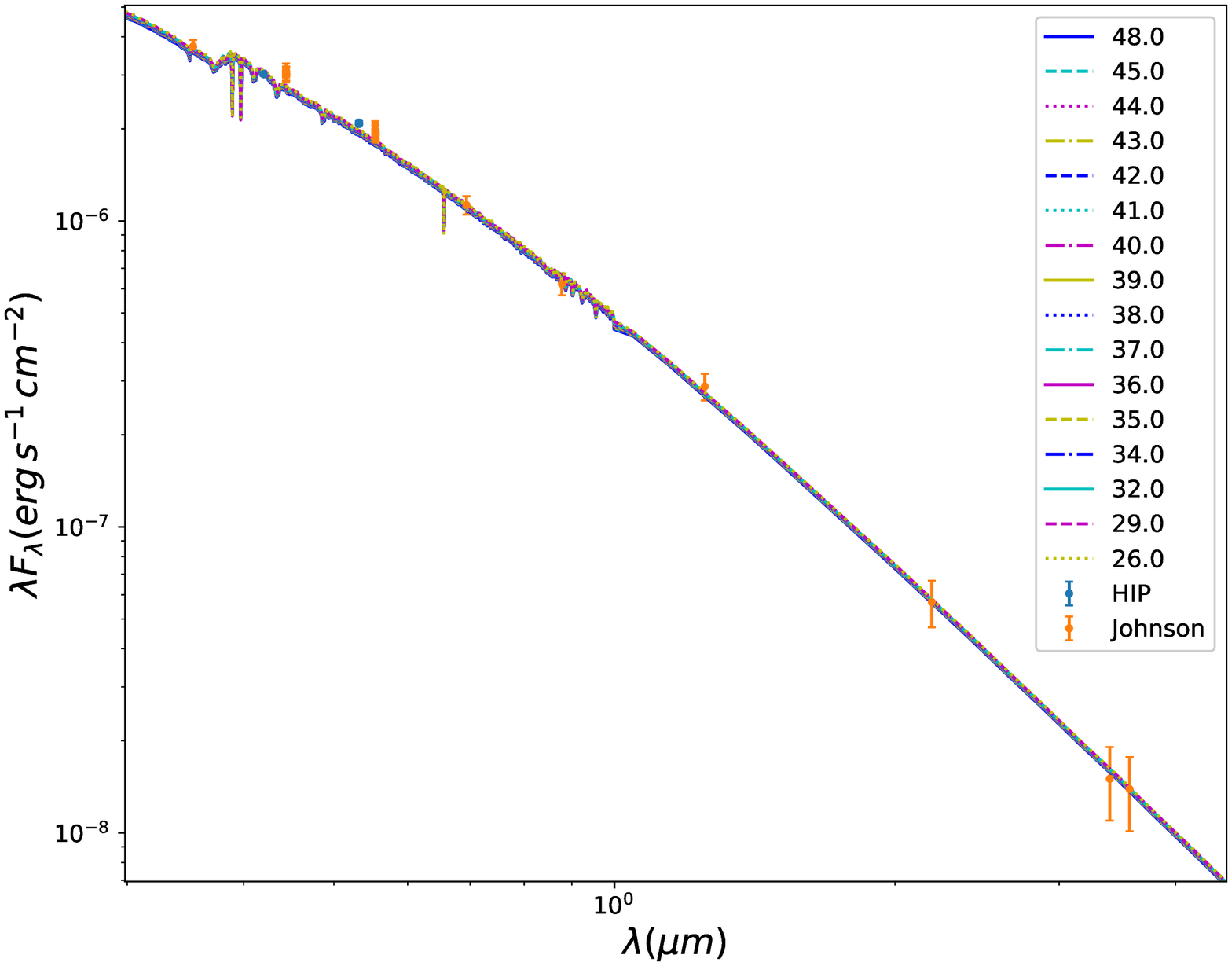}{0.4\textwidth}{(b)}}
    \caption{Diskless HDUST model for the primary of \del Sco using the parameters in Table \ref{tab:Best_BEMCEE_params} plotted with (a) the IUE UV spectra and (b) the photometry points from Vizier, which were used in BEMCEE as input to create the result of Figure \ref{fig:Best_BEMCEE_result}. The legend in each frame contains the inclinations of the HDUST model, and in (a) also contains the file names (in the upper left corner) of the FITS files obtained, which contain the IUE data, and in (b) indicates which filter system with which the photometric points were obtained.}
    \label{fig:diskless_HDUST_UV_SED}
\end{figure*}

A diskless model of the primary star of \del Sco was computed with HDUST using, as input, the parameters of $M$, $W$, $L$, and $R_p$ from Table \ref{tab:Best_BEMCEE_params}. The possible inclination range of \del Sco from Table \ref{tab:Best_BEMCEE_params} is listed between \ang{34} and \ang{45}. Since \del Sco is not rotating as rapidly as most Be stars ($W\, =\, 0.36$), there may not be much difference in the diskless model with a change in inclination, due to gravity darkening not having as strong of an effect as it would in a more rapidly rotating star \citep{McGill2013}. To see the effect of inclination on the diskless model of \del Sco, we computed the model for inclinations in the range of \ang{34} to \ang{45}  with steps of \ang{1}, and also extended our analysis to include angles of \ang{26}, \ang{29}, \ang{32}, and \ang{48} to better understand the effect of inclination as further explained below.

\begin{figure}[htb!]
    \plotone{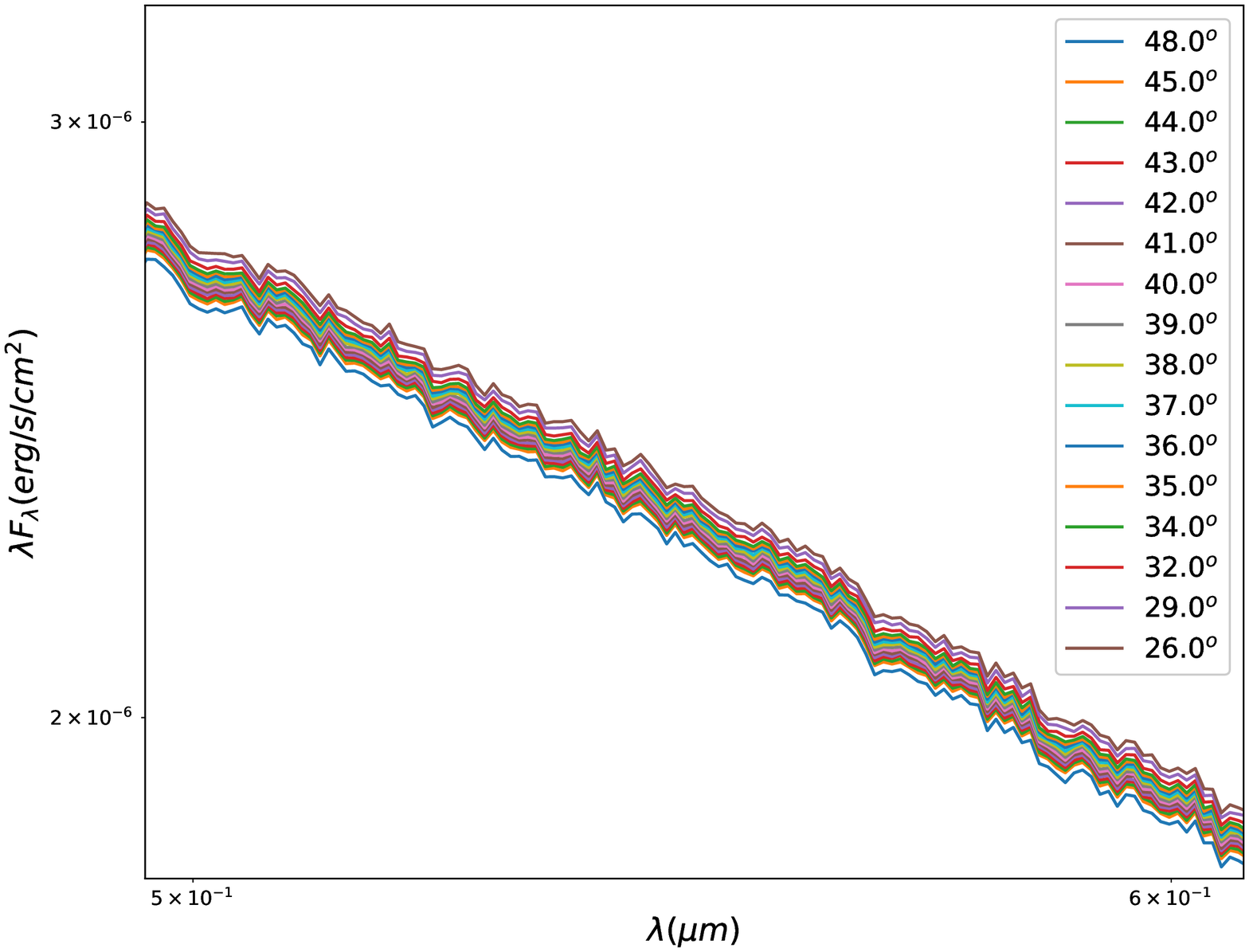}
    \caption{Diskless HDUST model for the primary of \del Sco zoomed in to the V-band wavelength regime ($\approx 0.5\text{ to }0.6 \, \rm \mu m$). The legend shows which colour corresponds to which inclination angle at which each model was computed.}
    \label{fig:diskless_sed_zoom}
\end{figure}

Figure \ref{fig:diskless_HDUST_UV_SED} shows the result of this diskless model plotted over (a) the IUE UV data and (b) the SED photometry points that were used as input into BEMCEE. Clearly, as expected, the change of inclination angles here has negligible effect on the diskless SED of \del Sco. Figure \ref{fig:diskless_sed_zoom} shows the diskless model in the V-band wavelength range only ($\approx 0.5\text{ to }0.6 \, \rm \mu m$) with the legend showing the colour of line that corresponds to which inclination angle the models were computed for. We calculate that the difference in V magnitude between the model at \ang{48} and \ang{26} is 0.04, and that the V magnitude scales linearly with increasing inclination angle. Thus the tight constraint determined from BEMCEE is most likely due to our use of the prior of $\rm \ang{38}\,\pm\,\ang{5}\,$. Given the good fit of our diskless model to our diskless data points as shown in Figure \ref{fig:diskless_HDUST_UV_SED}, we have confidence in the stellar parameters in Table \ref{tab:Best_BEMCEE_params}. To account for the invariance in the diskless model with inclination angle, we will expand our computed inclination angles for our models containing a disk beyond the bounds that have been determined from BEMCEE.

\subsection{Disk Models}
\label{models}

\begin{figure*}[htb!]
    \plotone{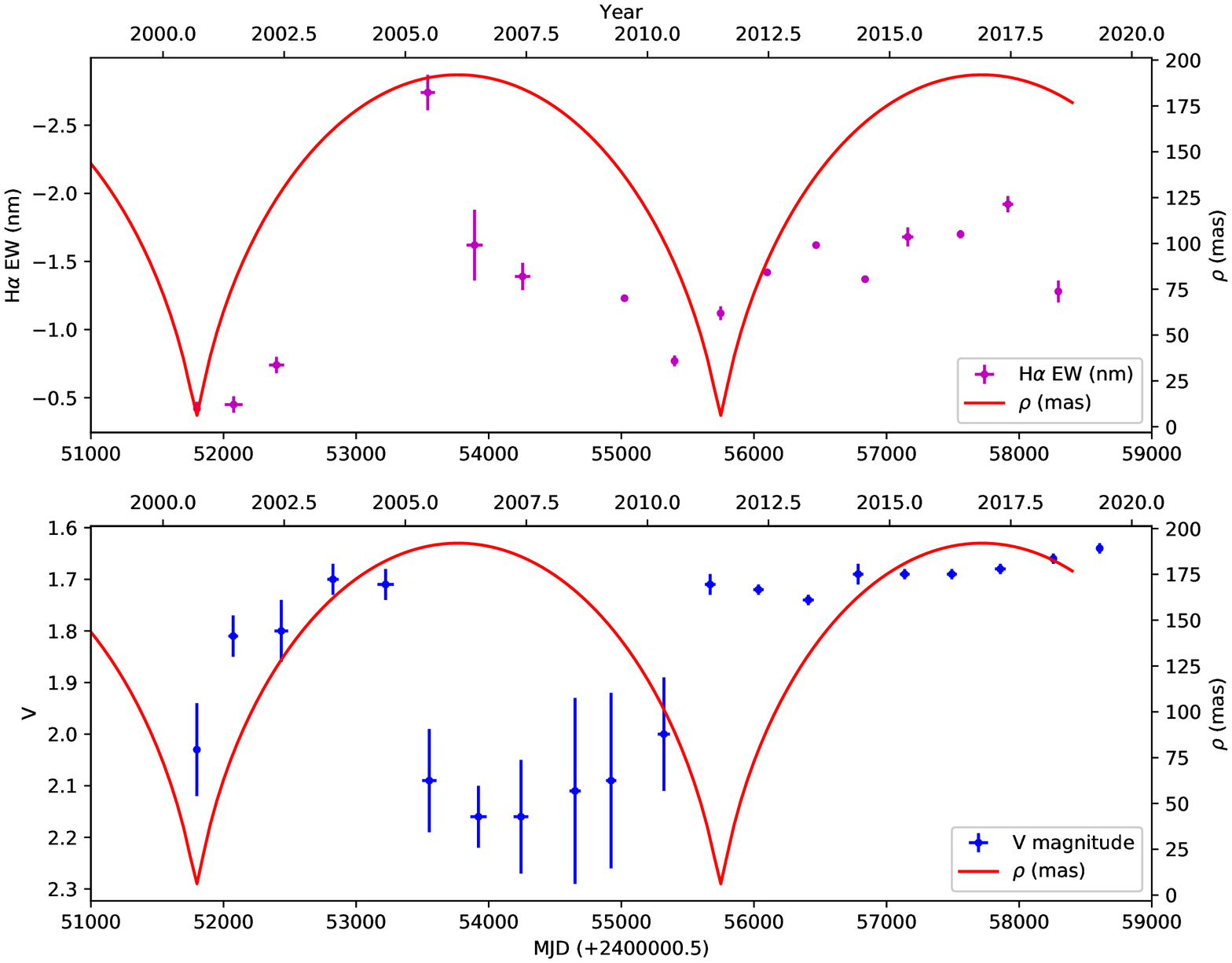}
    \caption{Similar to Figure \ref{fig:V_EW_plot} but with V magnitude and \Halpha EW, as well as their observation dates, averaged over each year. The error bars show the $1\sigma$ deviation of the data from the yearly average. As in Figure \ref{fig:V_EW_plot}, \Halpha EW is in the top plot in purple, V magnitude is in the bottom plot in blue, and the red line, common to both plots, is the separation between the primary and secondary stars. In the top and bottom plots, the left axis is the scale for the \Halpha EW  and  V  magnitude,  respectively,  while  the  right  axis  in  both  plots  is  the  scale  for  the  separation  of  the  two  binary components in mas.}
    \label{fig:V_Ha_yearly_avg}
\end{figure*}

To model the long-term trends of \del Sco, we averaged each year of data from Figure \ref{fig:V_EW_plot} beginning in 2000. These averages are displayed in Figure \ref{fig:V_Ha_yearly_avg}, a similar plot to Figure \ref{fig:V_EW_plot}, except we have plotted the yearly average of our V magnitude values, \Halpha EW, as well as the observation dates. The error bars indicate $1\sigma$ deviation of the data from the yearly average. The error bars also provide target ranges for our models of \del Sco. 

\begin{figure*}[hbp!]
    \gridline{\fig{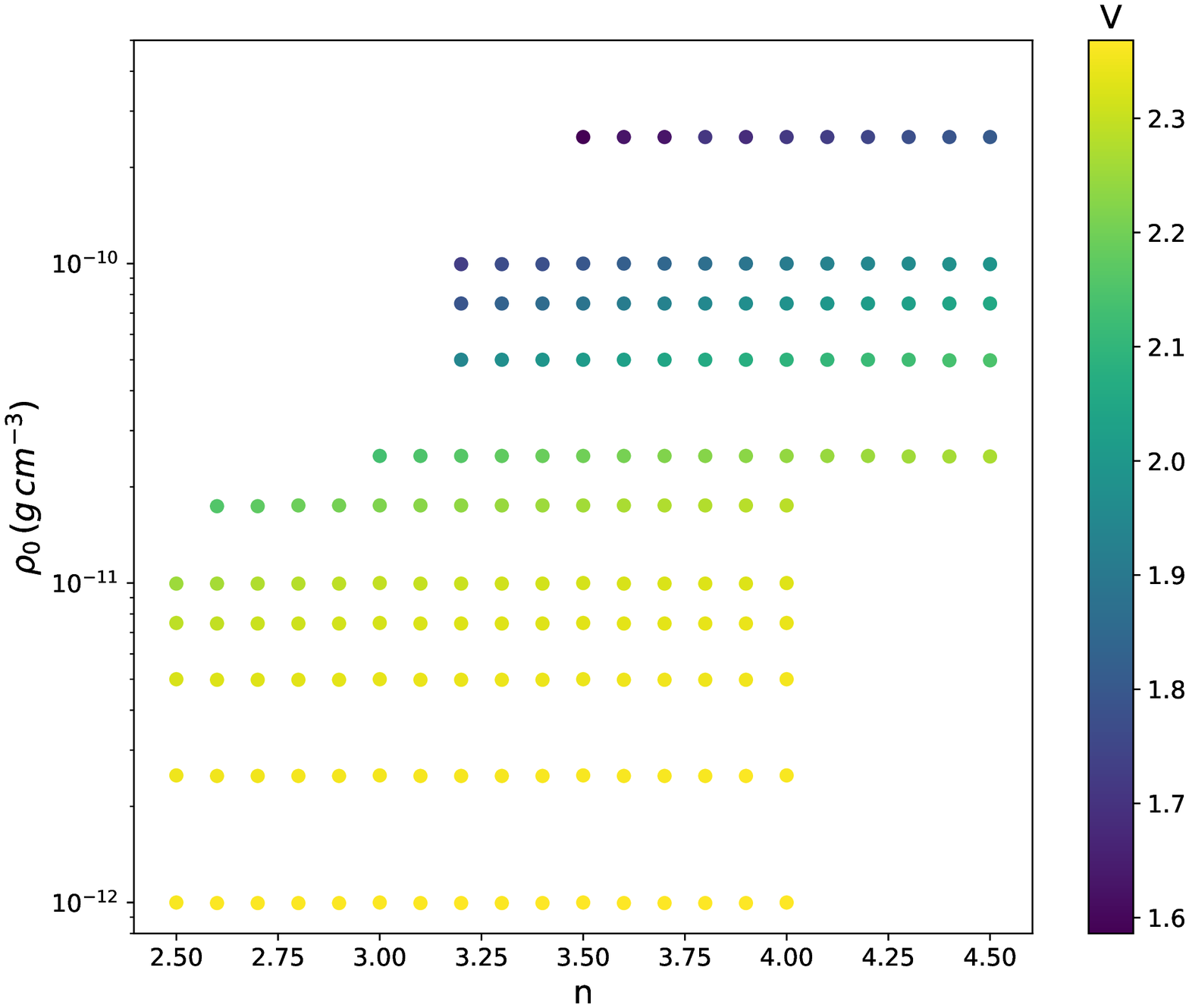}{0.4\textwidth}{(a)}
              \fig{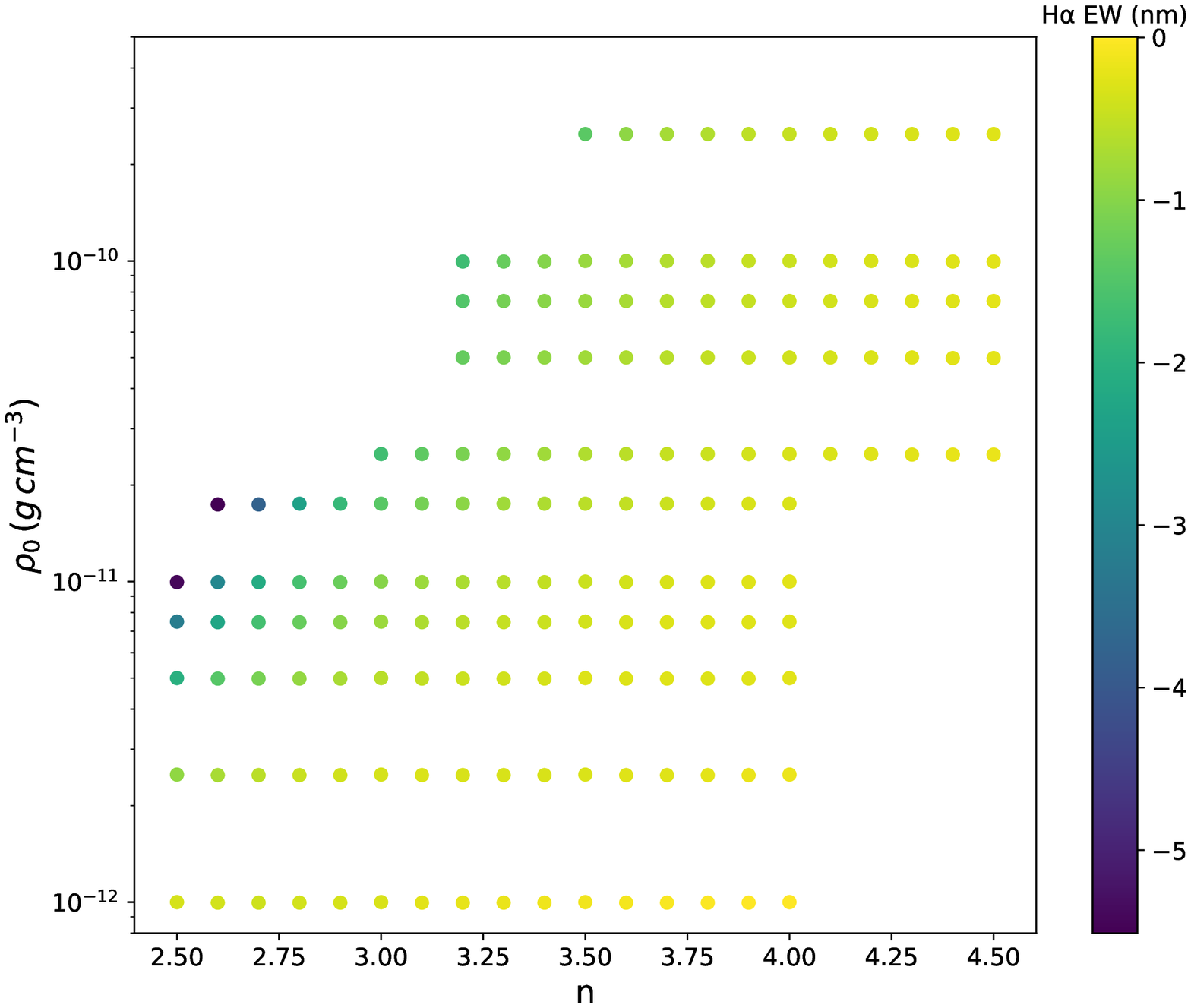}{0.4\textwidth}{(b)}}
    \caption{Complete grid of $n$ and $\rho_0$ values for the HDUST models of \del Sco. (a) is coloured by computed V magnitudes at an inclination of \ang{38}, and (b) is coloured according to calculated \Halpha EW values also at \ang{38}, the values of which are indicated in the colour bars on the right sides of both plots.}
    \label{fig:n_rho_grid}
\end{figure*}

In our modelling of \del Sco we vary the two variables $n$ and $\rho_0$ from Equation \ref{eq:volume_density}. We also vary the disk radius of each model, to ensure we capture all of the \Halpha emitting area of the disk. The combinations of $n$ and $\rho_0$ that were used can be seen in Figure \ref{fig:n_rho_grid}, where we have plotted the entire computational grid, with the colours corresponding to (a) calculated V magnitude values and (b) \Halpha EW values, at an inclination of \ang{38}. We computed our models for the range of inclination angles found for the diskless model of Figure \ref{fig:diskless_HDUST_UV_SED}, and for the additional inclinations of \ang{17}, \ang{20}, and \ang{23}. These three extra inclination angles were added because \cite{che2012imaging} found their best model of \del Sco to be at \ang{25}, and \cite{Arcos2017} found \ang{20} to be their best-fit model. Thus we wanted to investigate these inclinations for \del Sco for completeness. 

The trends of \Halpha and V magnitude seen in Figure \ref{fig:n_rho_grid} are also observed for the other inclination angles. If we keep $\rho_0$ constant but increase $n$, the EW will decrease while V remains fairly constant, and if we hold $n$ constant and increase $\rho_0$, then V will increase appreciably while the EW will not increase so much. These patterns in the calculated values allow us to determine an appropriate computational grid for values of $n$ and $\rho_0$. Once our models fall out of our acceptable range of values for \Halpha EW and V magnitude, we need not compute past this point because the observed trends tell us the calculated quantities will never reach our target values. This is the reason for the unevenness of our grid. That is, for values of $\rho_0$ on the order of $10^{-12}$ we stop computing at $n\, =\, 4$, while for higher values of $\rho_0$ we extend up to $n\, =\, 4.5$, but don't go lower than $n\, =\, 3$.

To compare our models to our yearly averaged data, we create, for each year, an averaged \Halpha spectrum from all of our observed spectra from a given year. The observations are resampled across 3200 evenly spaced points between $\pm 800$ km $\rm s^{-1}$ wavelength Doppler shift from the \Halpha line center, and an average is taken at each of these points to produce an averaged \Halpha spectrum. We then select only those models that have an \Halpha EW and V magnitude within the ranges of the  1$\sigma$ error bars for each year. We compare the models to the averaged spectra using the same figure-of-merit value, $\mathcal{F}$, as used in \cite{Jones2017}, which has the form,
\begin{equation}
    \mathcal{F} = \frac{1}{N}\sum_i w_i\frac{|F_i^{\rm obs}-F_i^{\rm mod}|}{F_i^{\rm obs}},
\label{eq:F_formula}
\end{equation}
where the weights are given by,
\begin{equation}
  w_i = \bigg|\frac{F_i^{\rm obs}}{F_c^{\rm obs}}-1\bigg|.
\label{eq:F_weights}
\end{equation}

$F_i^{\rm mod}$ is the model flux at wavelength $i$, $F_i^{\rm obs}$ is the observed flux at wavelength $i$, $F_c^{\rm obs}$ is the observed continuum flux level, and the sum is over all $N$ wavelength values denoted by $i$. This $\mathcal{F}$ value uses the weights, $w_i$, to put more emphasis on the middle portion of the line since our models do not account for non-coherent electron scattering which means that the wings of the predicted lines may be too narrow. Our best-fit model is the model with the lowest $\mathcal{F}$ value. We also adopt our error range of the best-fit to include any other model with an $\mathcal{F}$ within 20\% of the best-fit model.

\begin{figure*}[htb!]
    \plotone{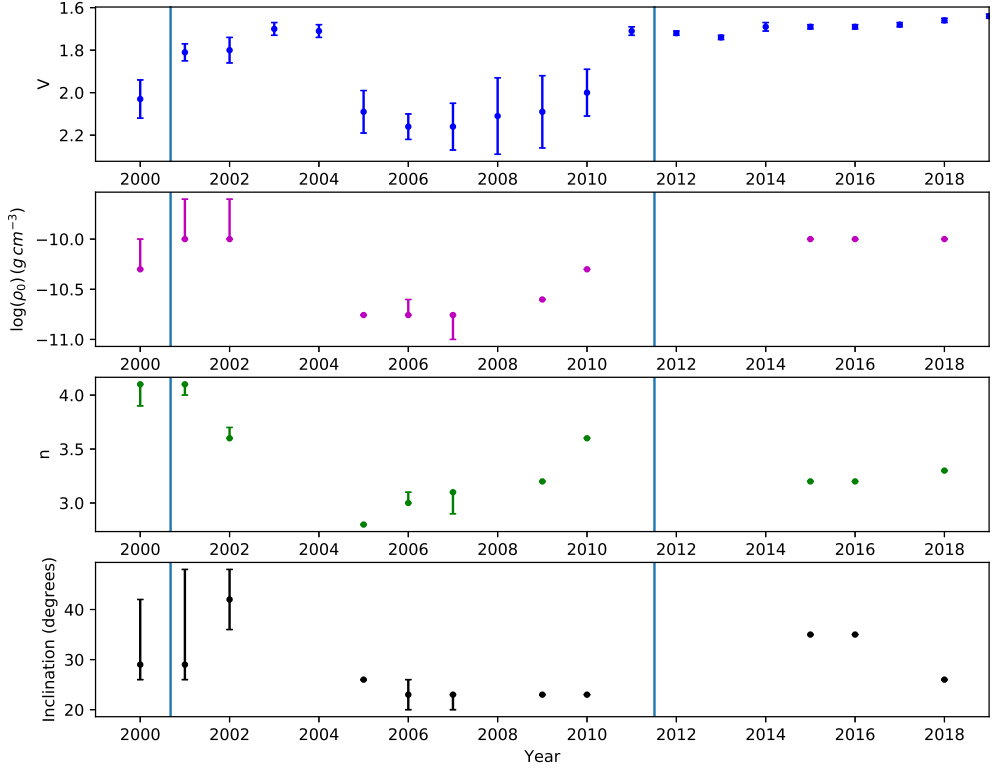}
    \caption{Results from our model fitting to the yearly averaged data with $1\sigma$ error. The top plot shows the V magnitude with its $1\sigma$ error. The other panels (top to bottom) are plots of the best-fitting $\rho_0$, $n$, and inclination for the disk of \del Sco. The error bars for the bottom 3 plots are the ranges for the various parameters from models that had an $\mathcal{F}$ within 20\% of our best-fitting model for each year. The vertical blue lines indicate the time of periastron in 2000 and 2011.}
    \label{fig:one_sigma_model_results}
\end{figure*}

The results from this fitting procedure are shown in Figure \ref{fig:one_sigma_model_results}, where we have plotted our V magnitude averages, and $1\sigma$ error, for comparison with the other three panels that show, in descending order, $\rho_0$, $n$, and inclination of our best-fit models, along with the error bars that show the ranges of the parameters from the models with an $\mathcal{F}$ within 20\% of our best-fitting model for each year. The vertical blue lines on each plot  indicate the time of periastron in 2000 and 2011.

Our best-fitting $\rho_0$ values reach a maximum of $1 \times 10^{-10} \, \rm g\, cm^{-3}$ in 2001, 2002, 2015, 2016, and 2018, and have a minimum value of $1.75 \times 10^{-11} \, \rm g\, cm^{-3}$ in 2005, 2006, and 2007. The $n$ values range from 2.8 to 4.1, and the best-fit inclination values range from \ang{23} to \ang{42}. By plotting the V magnitude along with these three parameters in Figure \ref{fig:one_sigma_model_results}, we can see that $\rho_0$ (and, to an extent, $n$ and inclination) seems to follow the same oscillation pattern as the observed V magnitude, reaching a maximum in the early 2000s before dropping down to a minimum during 2005 to 2007, and then increasing through 2009 to 2010, before finally reaching and sustaining a maximum in the following years.

\begin{figure*}[htb!]
    \plotone{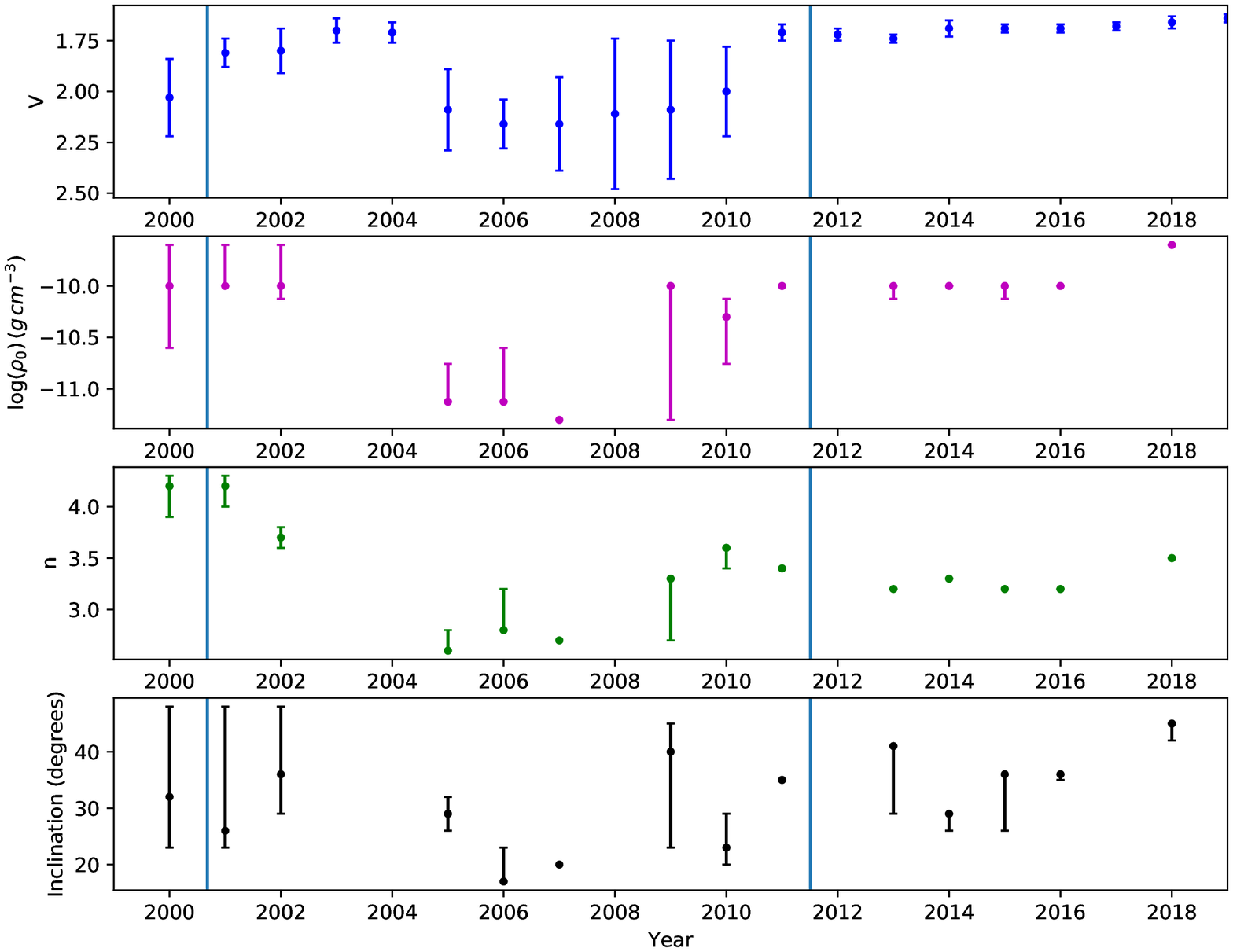}
    \caption{Results from our model fitting to the yearly averaged data with $2\sigma$ error. The top plot shows the V magnitude with its $2\sigma$ error. The plots of the other parameters are of the same format as Figure \ref{fig:one_sigma_model_results}. As previously, the vertical blue lines indicate the time of periastron in 2000 and 2011.}
    \label{fig:two_sigma_model_results}
\end{figure*}

These yearly averaged results from the $1\sigma$ constraints are restrictive and do not find HDUST models within the appropriate ranges of EW and V magnitude several of the years we investigated, including 2011 to 2014 and 2017, so we applied the same procedure as above to our averaged data with a $2\sigma$ error to see if the larger range of possible values finds suitable models in these years. These results are shown in Figure \ref{fig:two_sigma_model_results}, where we have the same format as in Figure \ref{fig:one_sigma_model_results}, except in the top plot where the V magnitude averages have error bars of $2\sigma$. Expanding our range of values gives us more years (including 2011, 2013, and 2014) with matching models, but results in larger errors on the best-fit parameters. We see similar limits to our best-fit parameters here as in the $1\sigma$ case: $\rho_0$ ranges from $2.5 \times 10^{-10} \, \rm g\, cm^{-3}$ in 2018 to $5 \times 10^{-12} \, \rm g\, cm^{-3}$ in 2007, $n$ ranges between 2.6 and 4.2, and inclination has values from \ang{17} to \ang{45}. There is also a similar oscillation pattern in the values of these parameters as in the $1\sigma$ case. It is also worth noting that different methods of weighting the \Halpha EW, V magnitude, and $\mathcal{F}$ to determine the best-fit models yield the same general long-term behaviour of \del Sco.

\begin{figure*}[p!]
\gridline{\fig{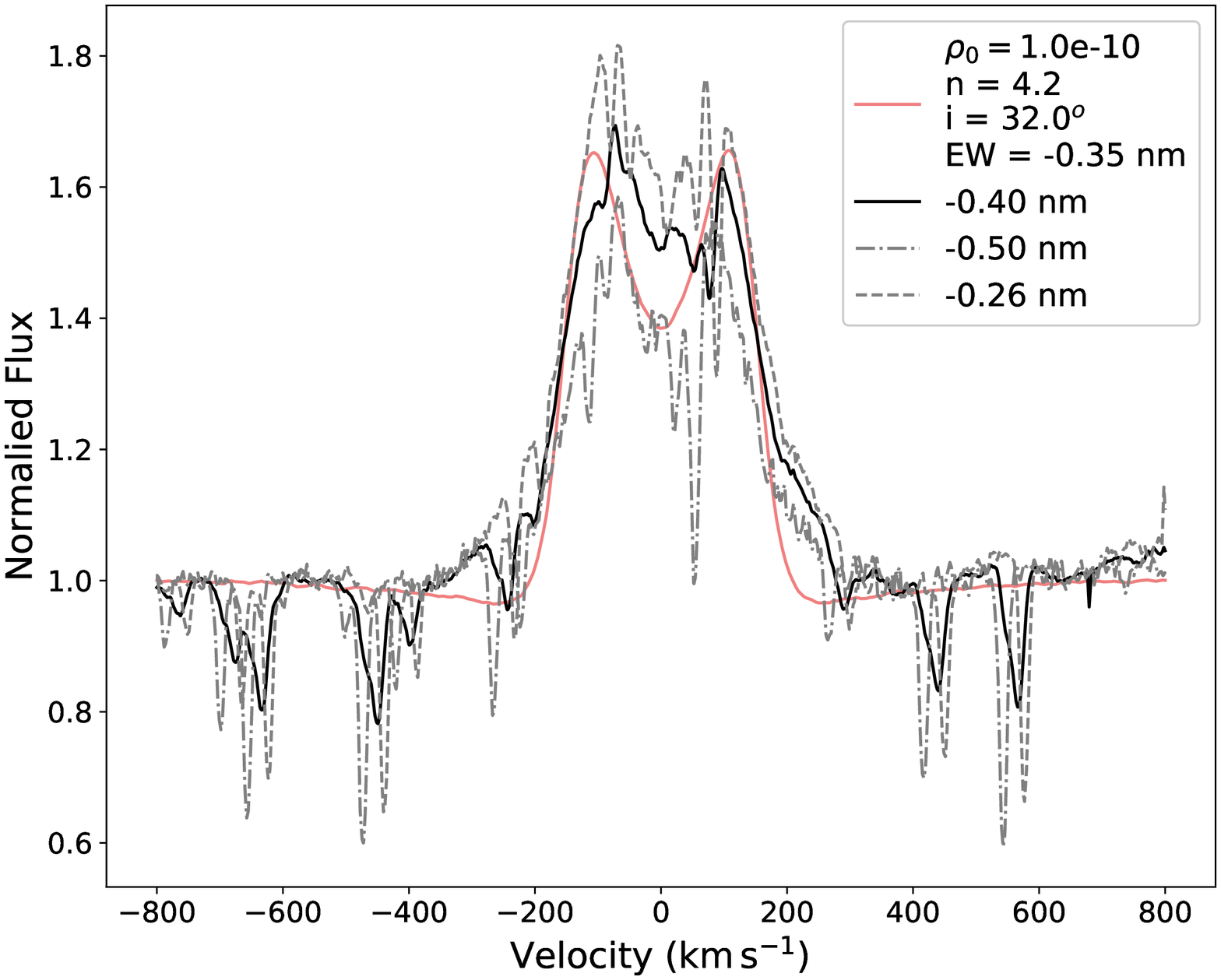}{0.3\textwidth}{(a)2000}
        \fig{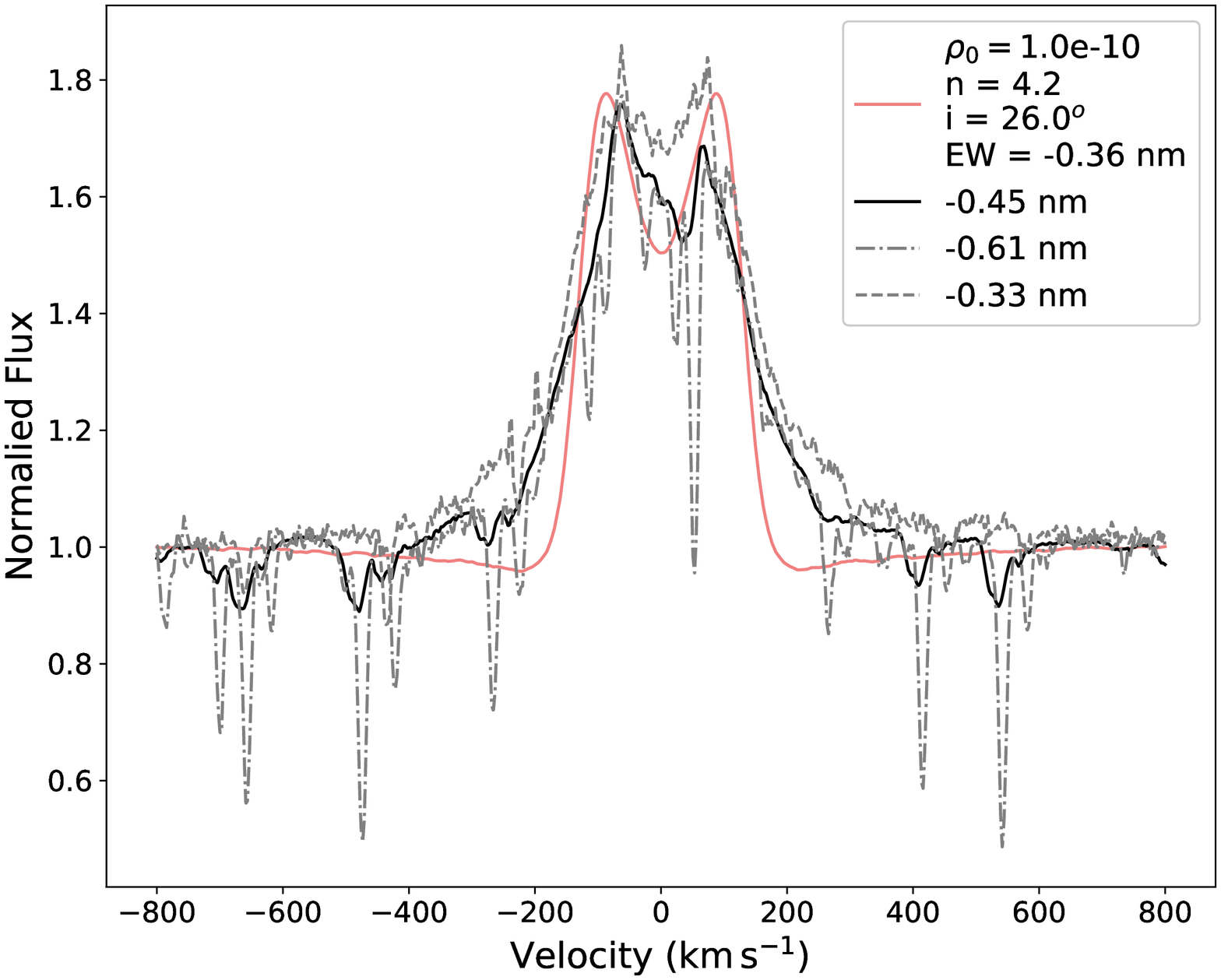}{0.3\textwidth}{(b)2001}
        \fig{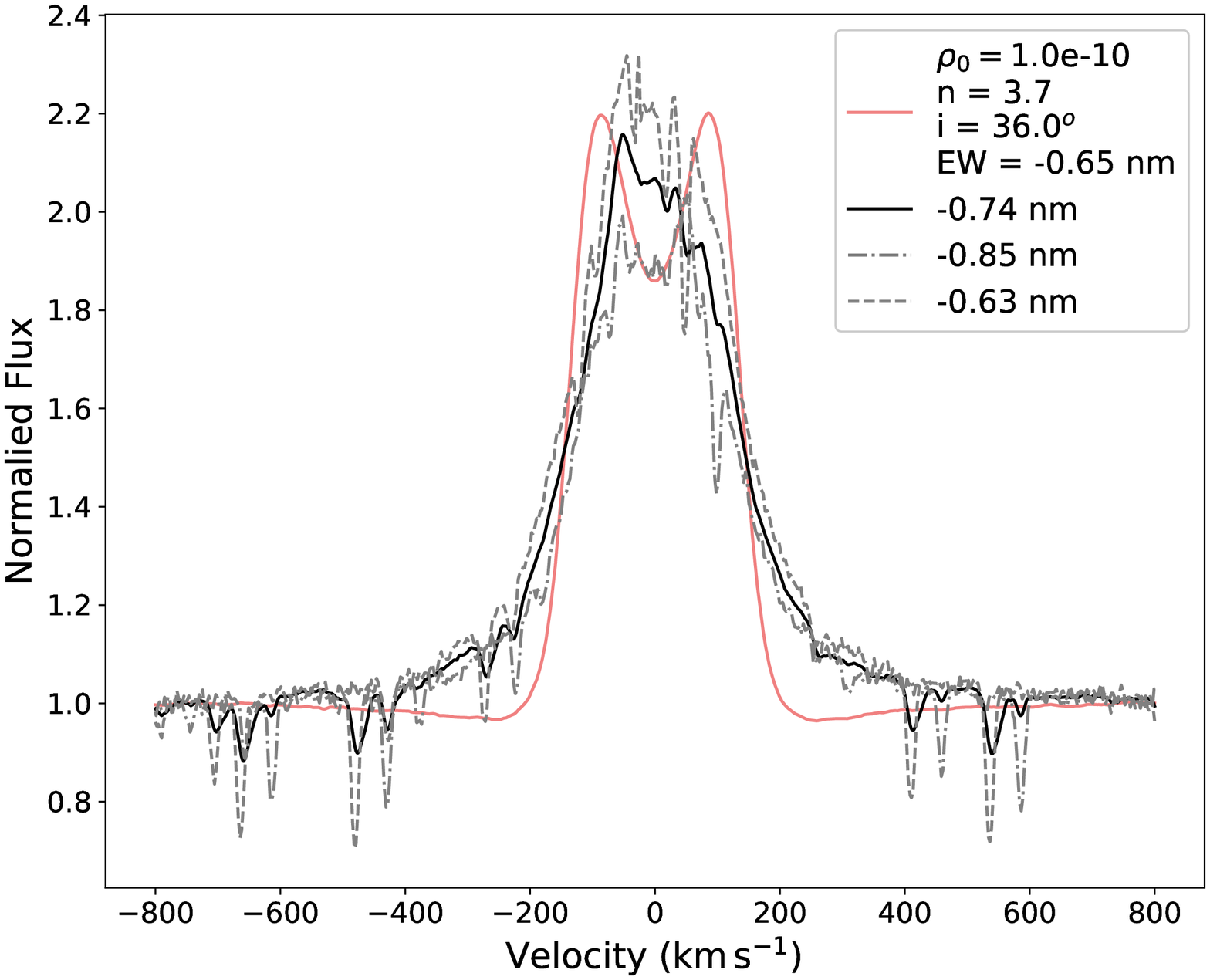}{0.3\textwidth}{(c)2002}}
\gridline{\fig{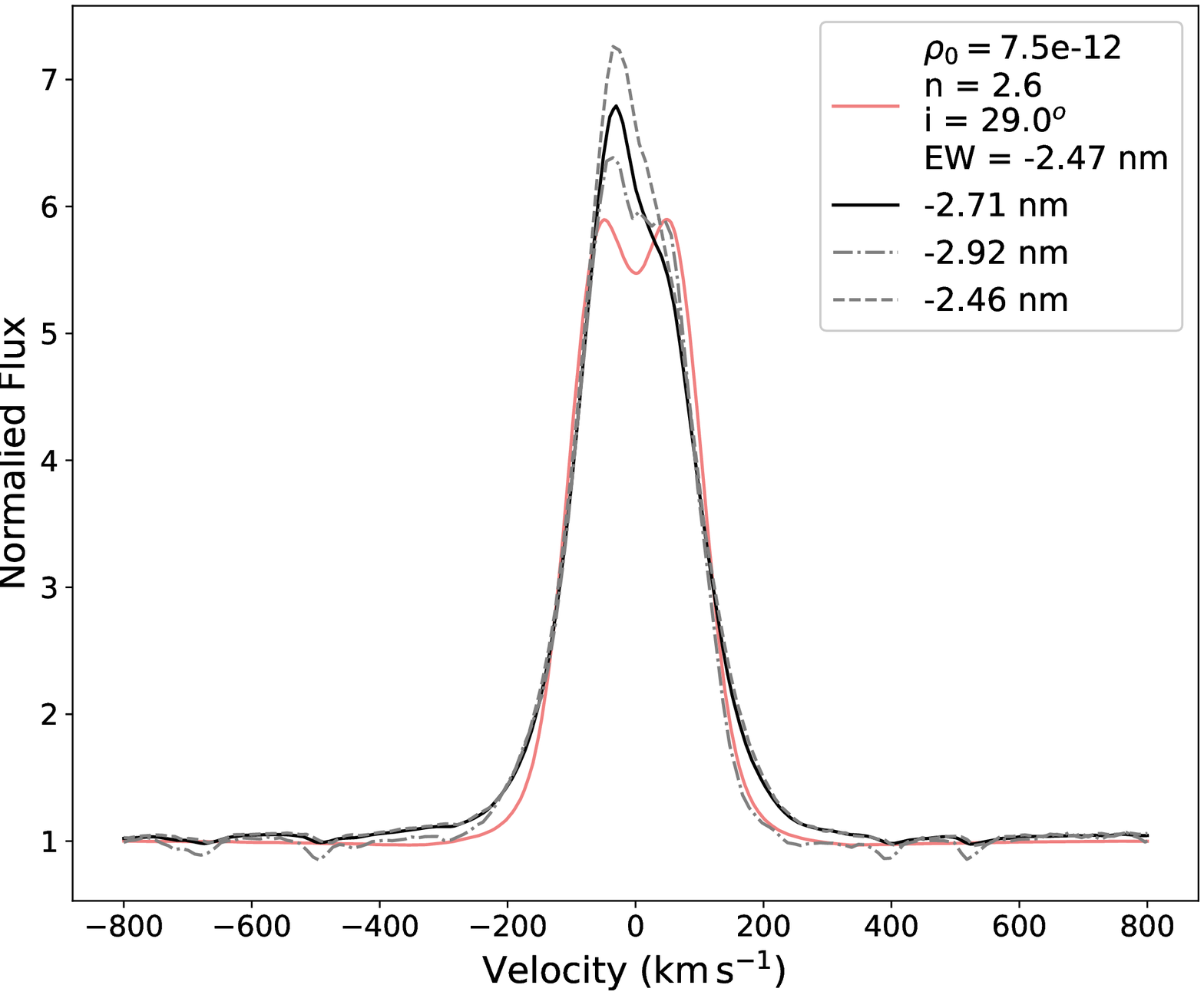}{0.3\textwidth}{(d)2005}
        \fig{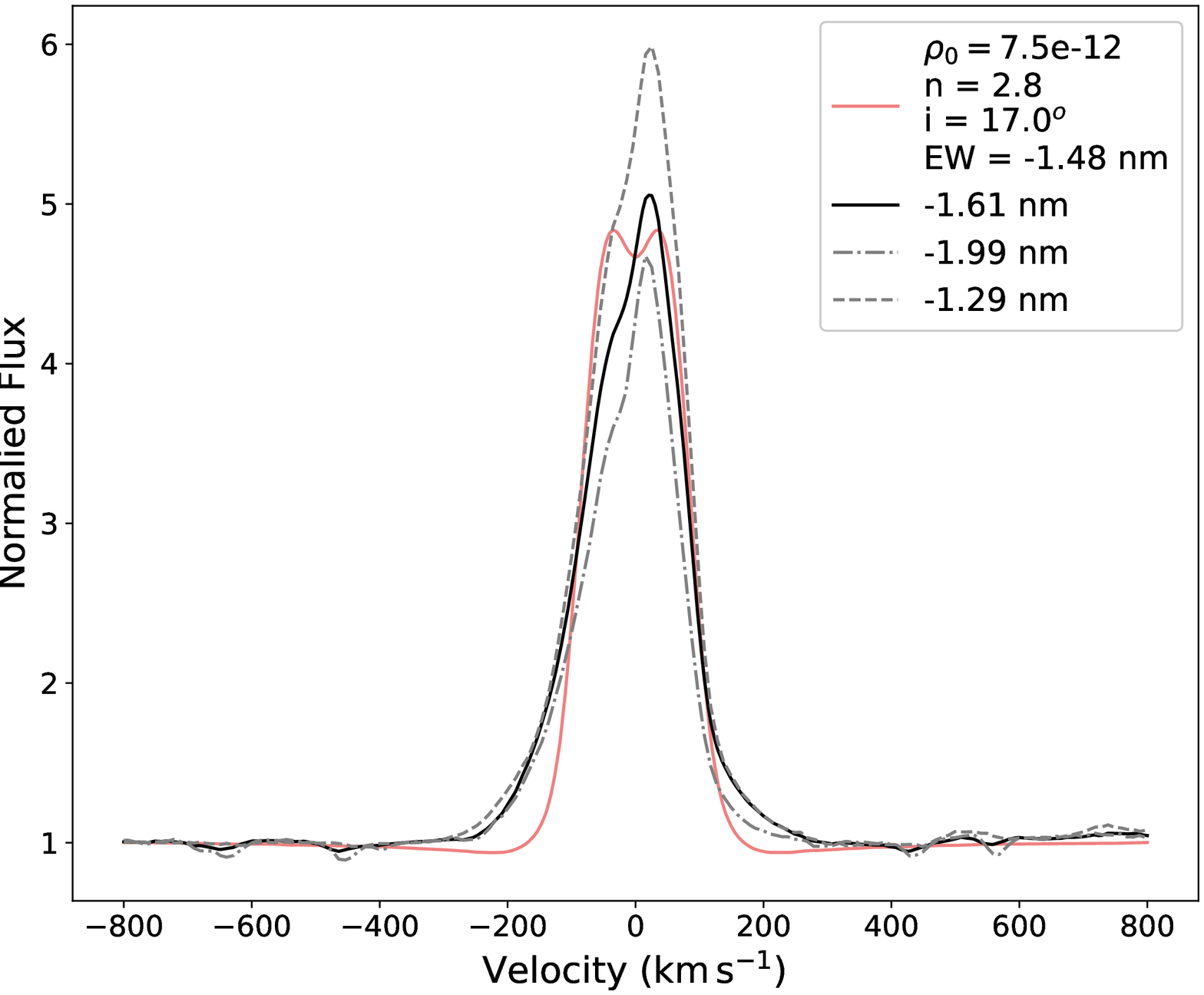}{0.3\textwidth}{(e)2006}
        \fig{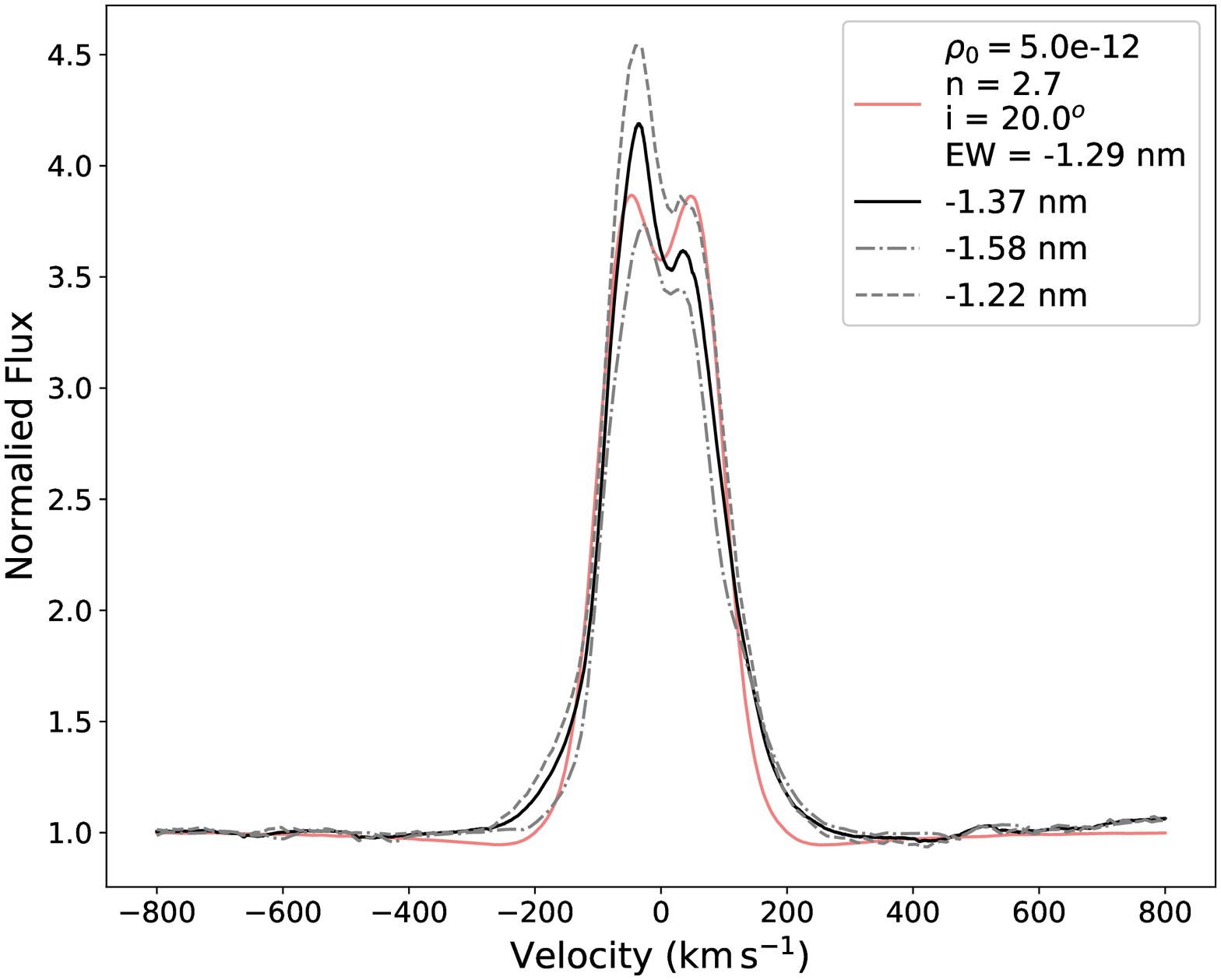}{0.3\textwidth}{(f)2007}}
\gridline{\fig{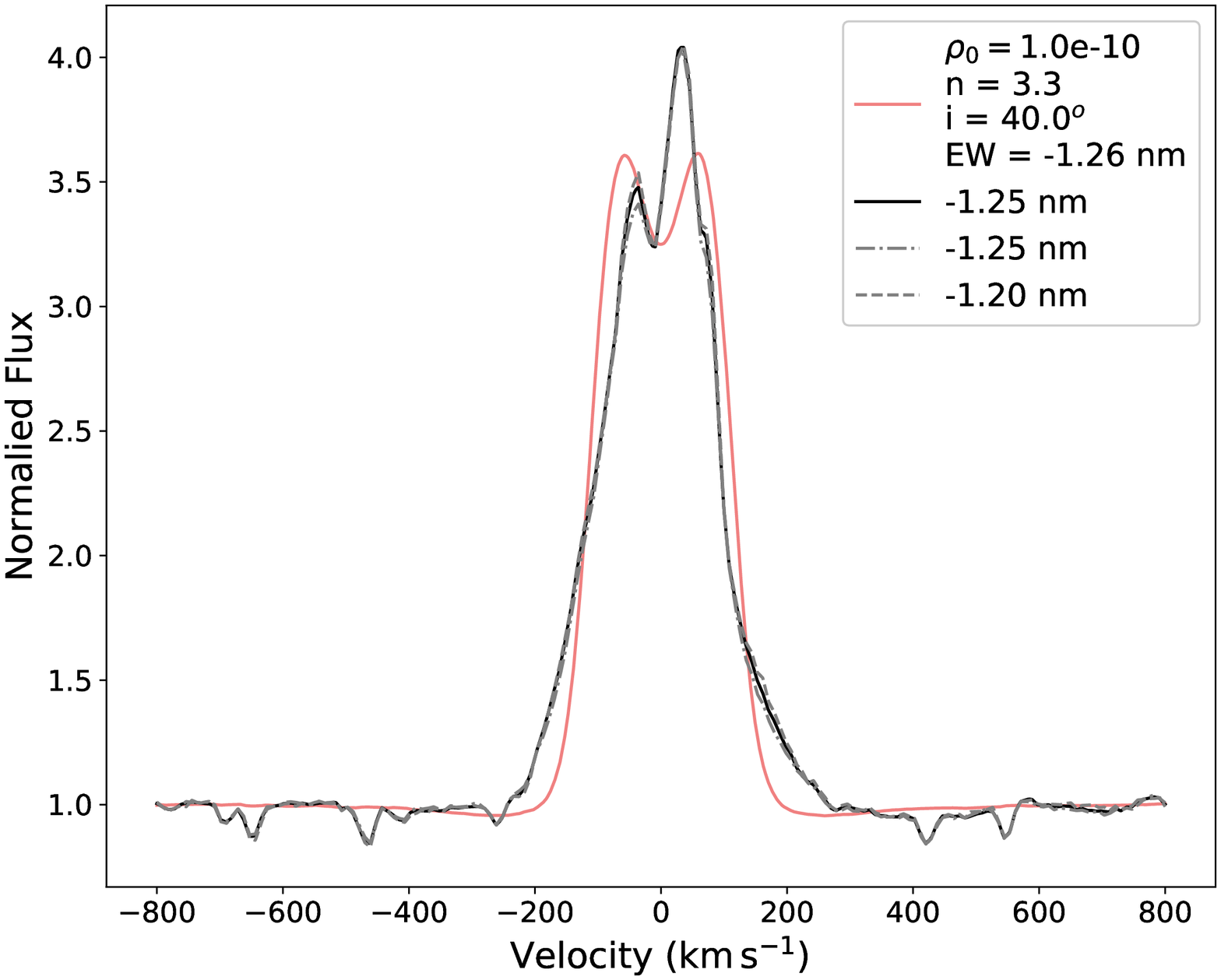}{0.23\textwidth}{(g)2009}
        \fig{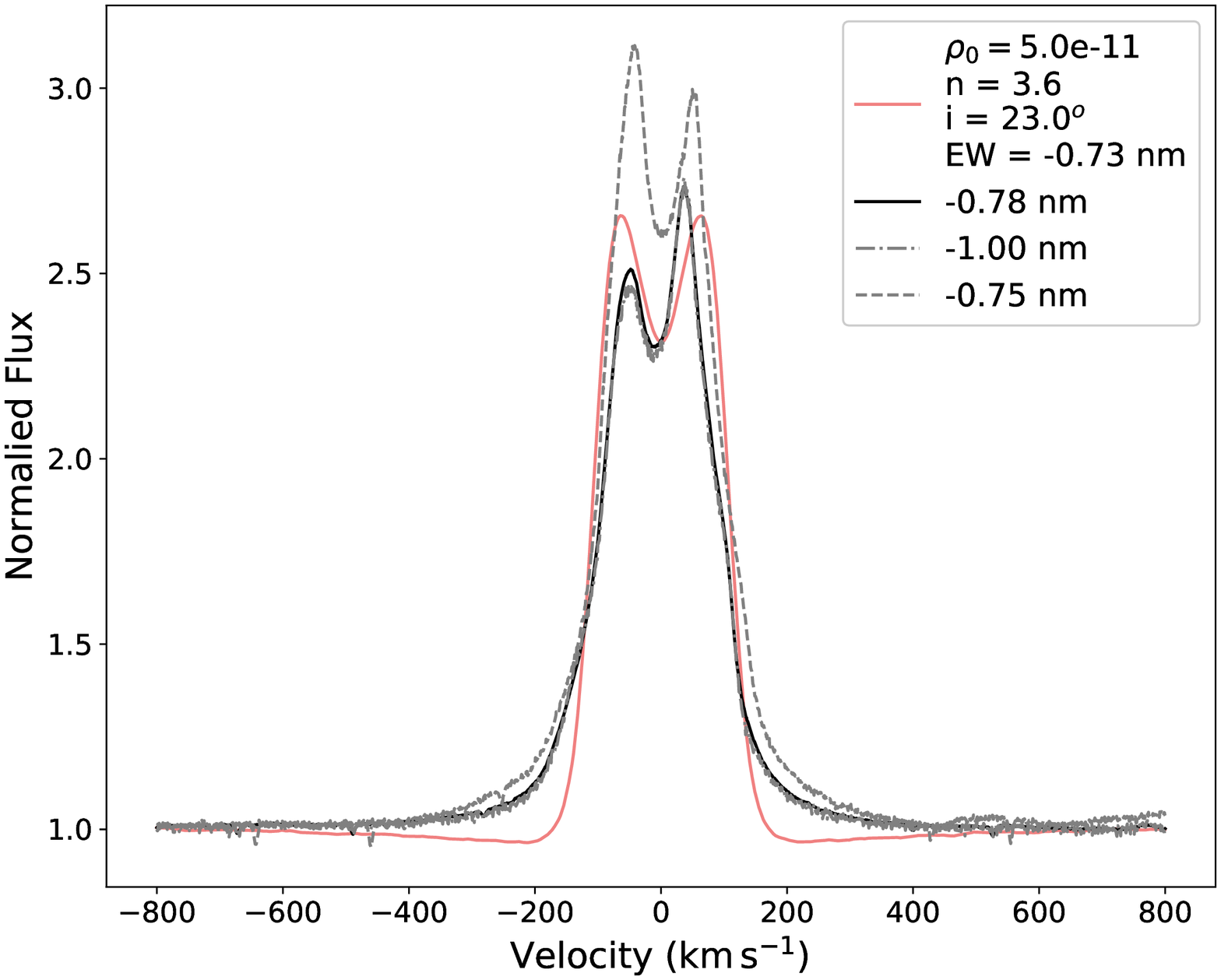}{0.23\textwidth}{(h)2010}
        \fig{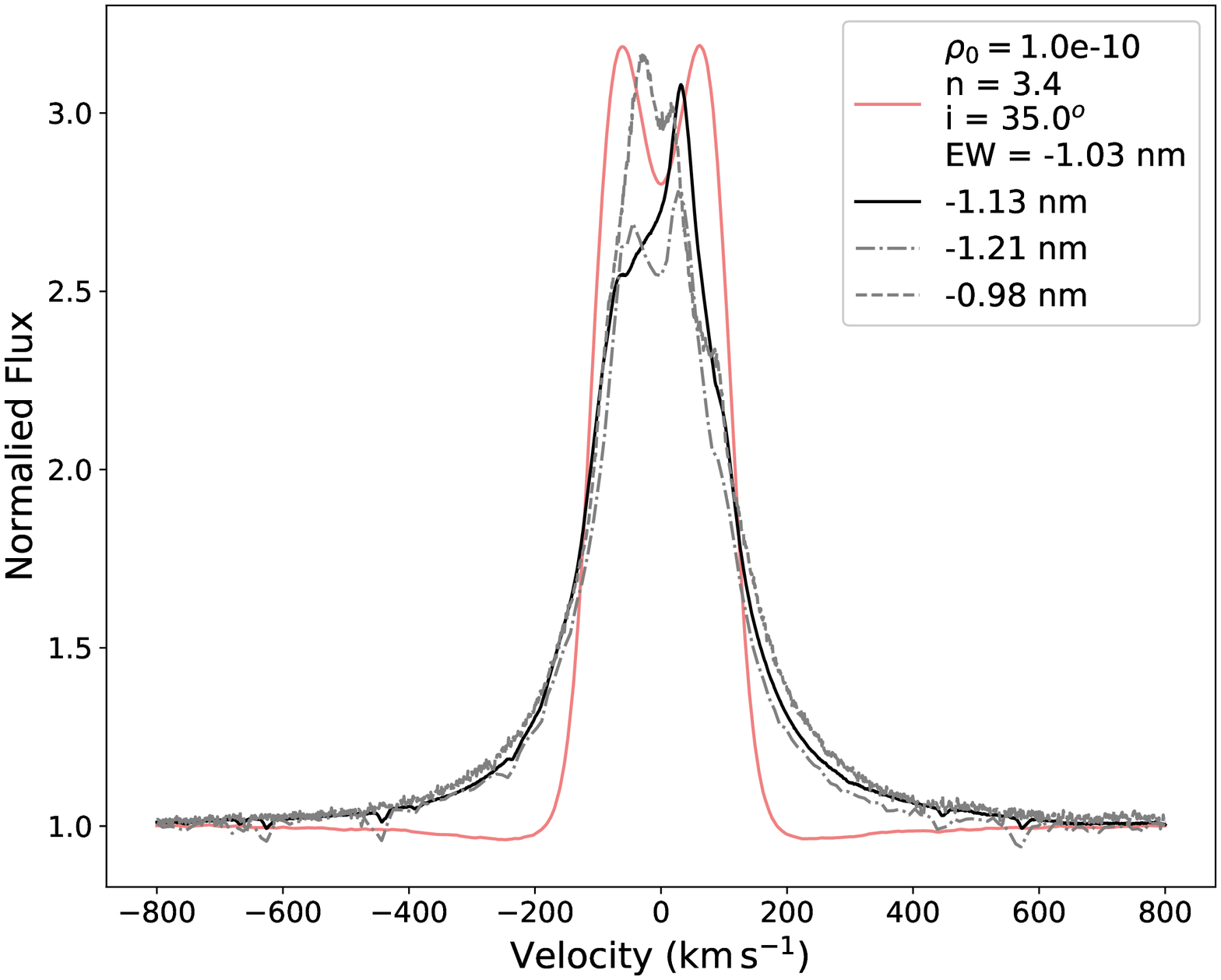}{0.23\textwidth}{(i)2011}
\fig{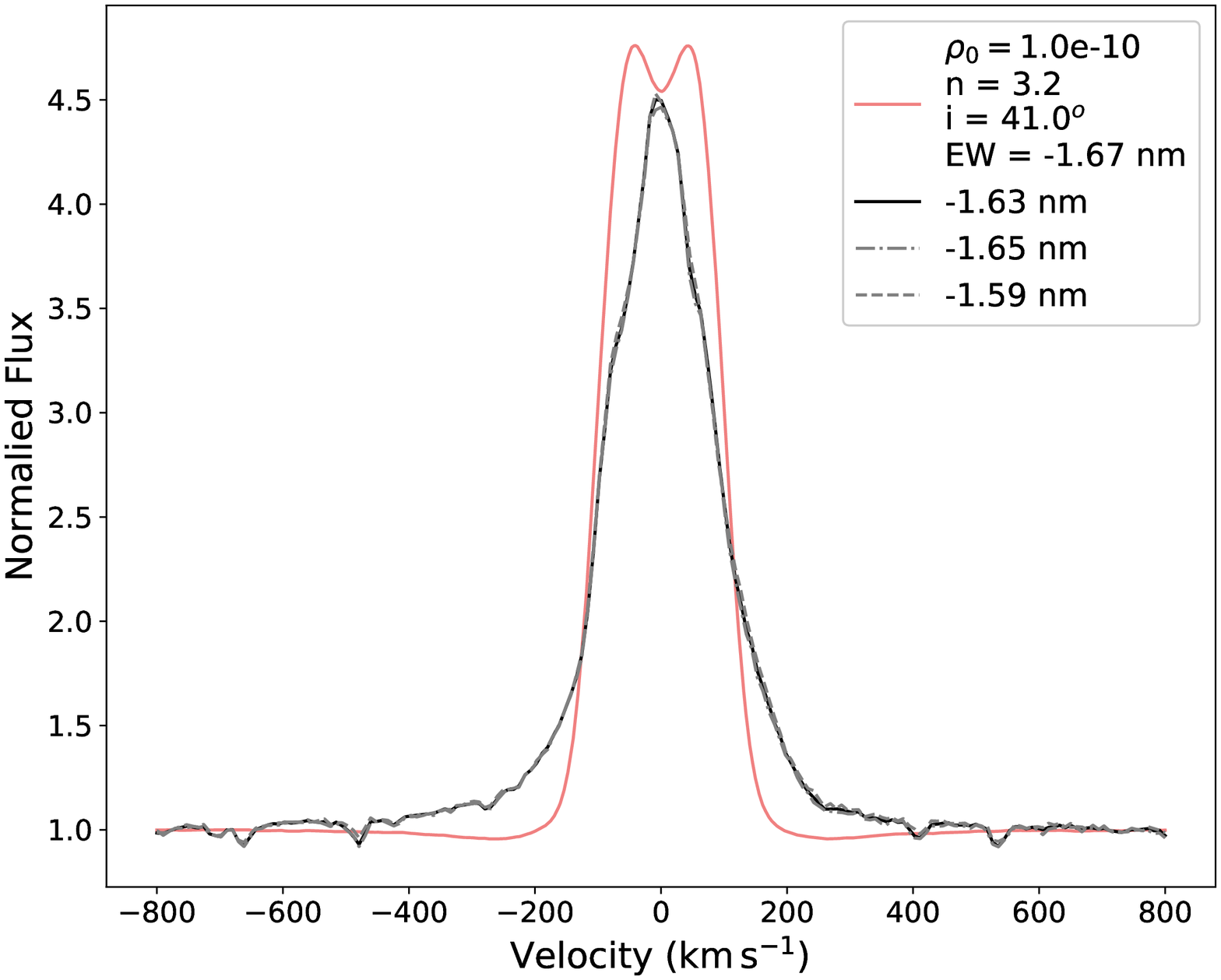}{0.23\textwidth}{(j)2013}}
\gridline{\fig{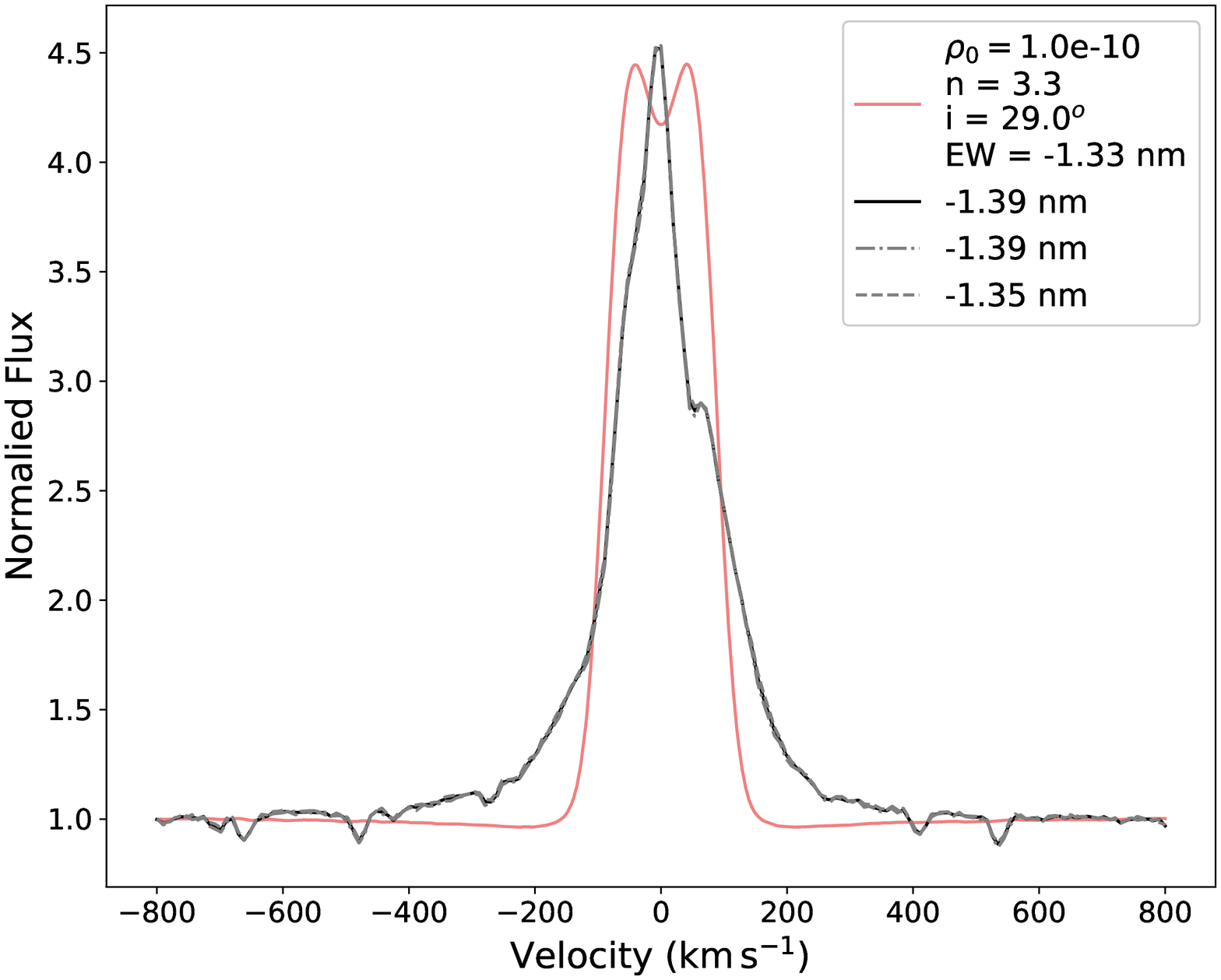}{0.23\textwidth}{(k)2014}
        \fig{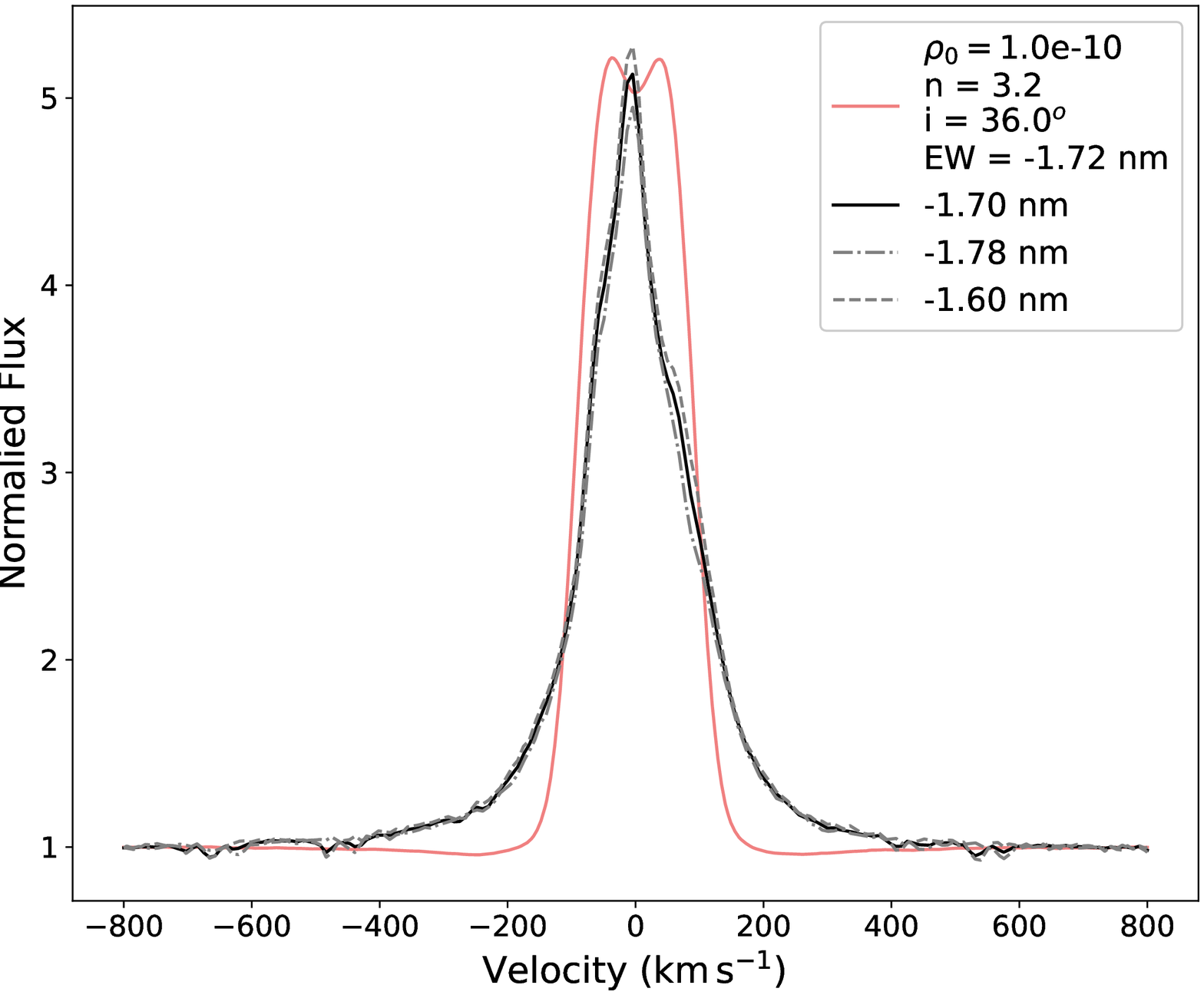}{0.23\textwidth}{(l)2015}
    \fig{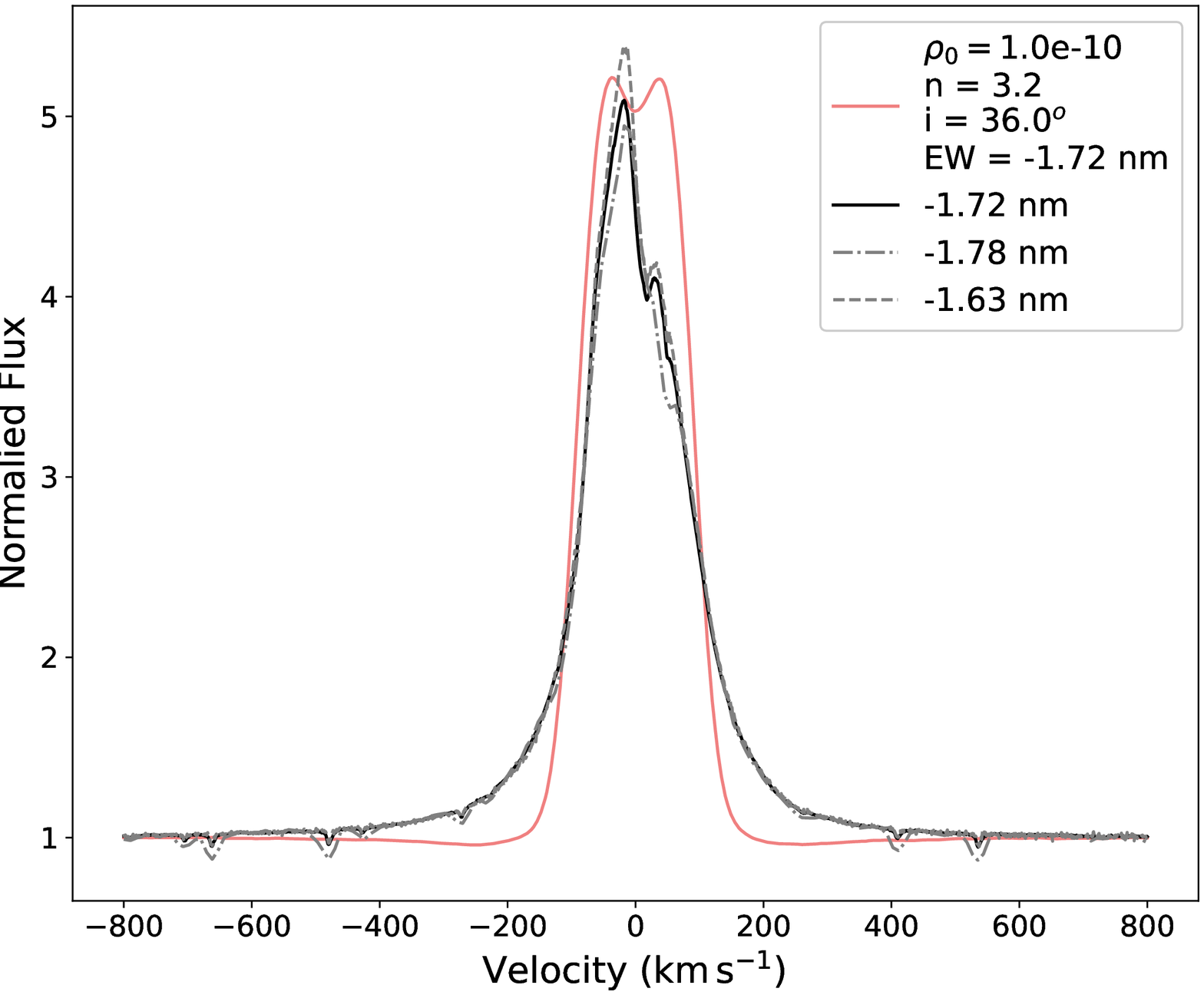}{0.23\textwidth}{(m)2016}
        \fig{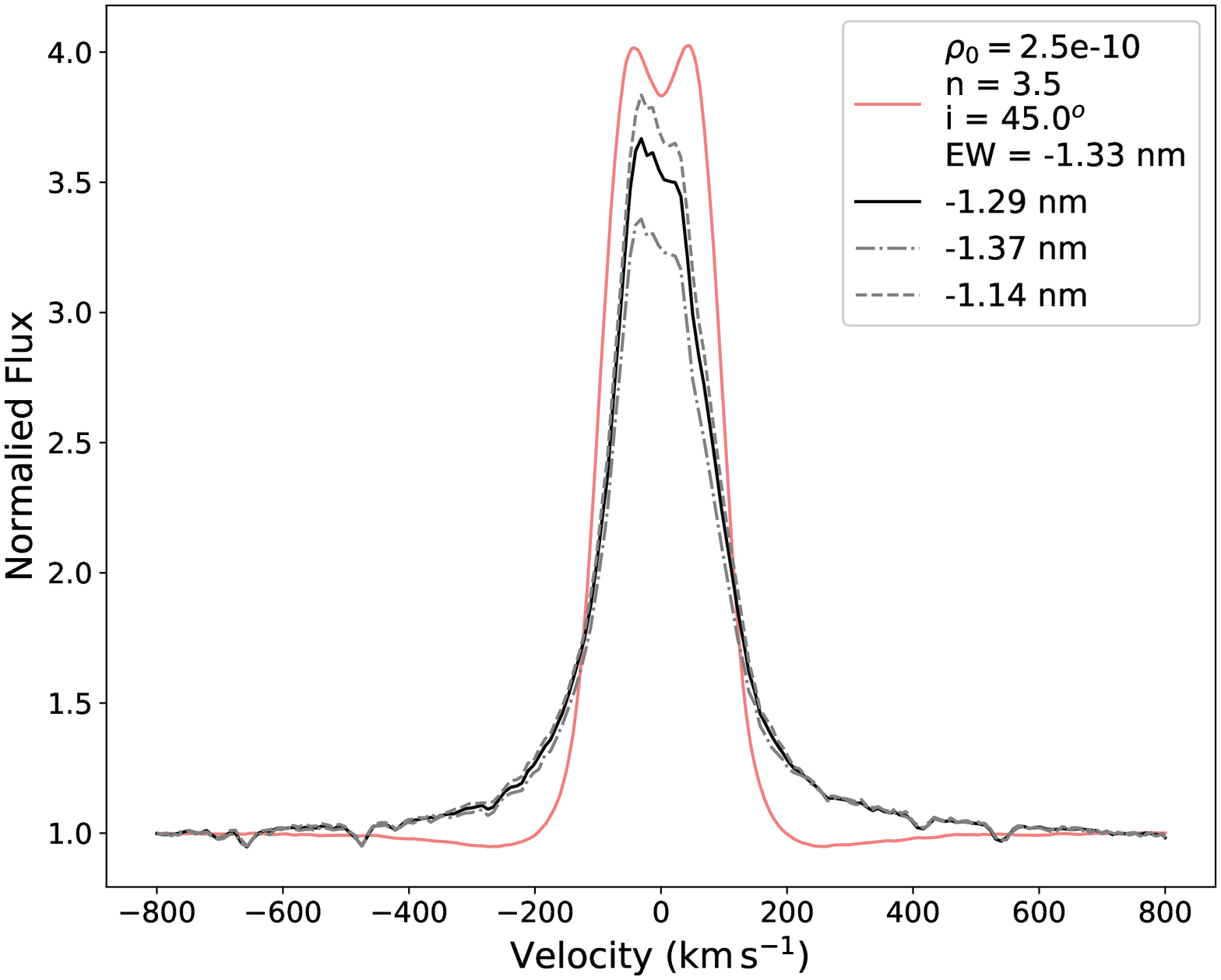}{0.23\textwidth}{(n)2018}}
    \caption{Best-fit plots from our model fitting to the yearly averaged data with $2\sigma$ error. In each plot, the red line is the best-fitting model determined by the $\mathcal{F}$ formula. The black line is the averaged observed spectra of each year, the grey dashed-dotted line is the spectra from the corresponding year with the smallest EW, and the grey dashed line is the spectra with the largest EW for the specified year. The year that each plot corresponds to is provided in the caption of each plot. The legend indicates the values of $\rho_0$, $n$, inclination, and EW (in nm) for the HDUST models, and states the EW for the average, maximum, and minimum observed spectra.}
    \label{fig:two_sigma_model_plots}
\end{figure*}

Figure \ref{fig:two_sigma_model_plots} displays plots of our best-fit \Halpha spectra produced using HDUST (red line), our averaged observed spectra (black line), the spectra with the smallest EW (grey, dashed-dotted line), and the spectra with the largest EW (grey, dashed line) for each year when we had a model fall into the appropriate $2\sigma$ range of \Halpha EW and V magnitude. The year for each plot is indicated in the sub-captions directly below each panel. Almost all of our best-fitting models do not match the observations well in the wings of the spectral line. This is both due to the fact that our $\mathcal{F}$ calculation favoured a model which matched the center of the line better than the wings, and that HDUST does not account for non-coherent electron scattering, as previously discussed. However, the HDUST models seem to provide a fair representation of the long-term changes of the \Halpha profile.

\begin{figure}[htb!]
    \plotone{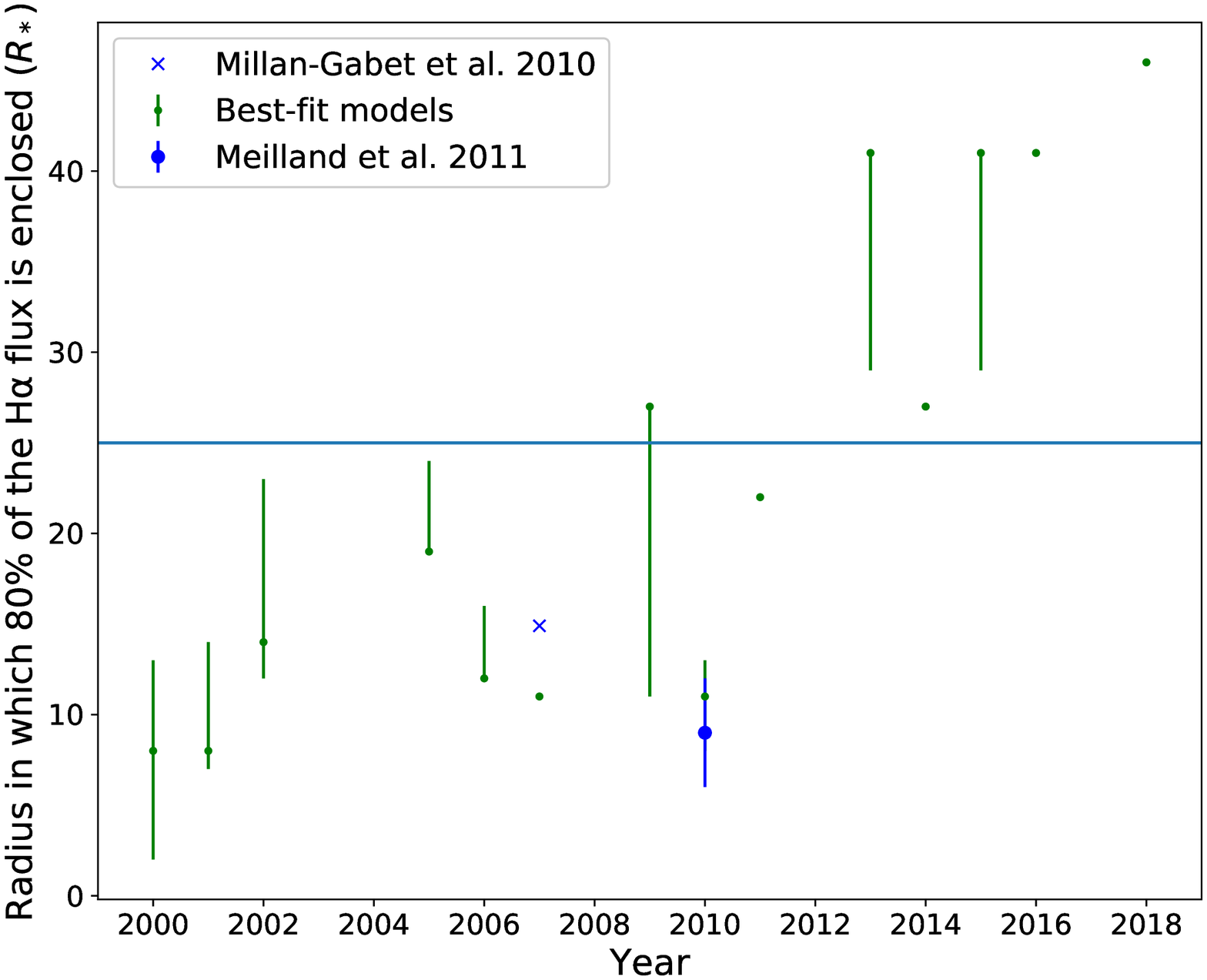}
    \caption{Size of the \Halpha emitting region of our models within which 80\% of the \Halpha emitted flux is enclosed. The error bars come from the same models that had an $\mathcal{F}$ within 20\% of our best-fitting models also used in Figure \ref{fig:two_sigma_model_results}. The size of the \Halpha emitting region from \cite{millan2010spectro} (blue x) and \cite{meilland2011binary} (blue circle) is also shown along with a blue line which indicates the periastron separation between the primary and secondary stars.}
    \label{fig:two_sigma_model_radii}
\end{figure}

We also explored the radial extent of the \Halpha region of our disk models for \del Sco. Figure \ref{fig:two_sigma_model_radii} shows the radius within which 80\% of the emitted \Halpha flux is contained for our best-fitting models. The error bars, as in Figure \ref{fig:two_sigma_model_results}, come from the models that had an $\mathcal{F}$ value within 20\% of the best-fitting models. It also shows two results from previous works of the size of the \Halpha emitting region from \cite{millan2010spectro} and \cite{meilland2011binary}, and has a blue line indicating the periastron distance between the primary and secondary stars of \del Sco. From this plot we see noticeable radial growth of the disk from 2000-2018. The disk started off small, around 10 $R_*$ early in the century, and grew, in our models, up to 46 $R_*$ by 2018. We also see that our results for 2007 and 2010 agree well with the sizes of $14.9\, R_*$ of \cite{millan2010spectro} and $9\pm3\, R_*$ of \cite{meilland2011binary}. It is important to recognize that these radii represent the size of the \Halpha emitting region of the disk, not the total physical extent of the disk.

\section{Discussion}
\label{sec:discussion}

Since the year 2000, \del Sco has gone through clear disk building and dissipating phases as shown in our V-band photometry in Figure \ref{fig:V_EW_plot}. The photometric behaviour of \del Sco from 2000 to 2011 looks very similar to that of $\omega$ CMa in \cite{Ghoreyshi2018}, which showed four periods of disk building and dissipation through V-band photometry. The V-band photometry of \del Sco from 2000 to 2011 closely resembles the repeating cycle of the light curve for $\omega$ CMa (figure 1 of \citealt{Ghoreyshi2018}).  From the observational evidence of \cite{miroshnichenko2001spectroscopic}, \del Sco showed its first clear evidence of a disk in 2000. This is confirmed by our collected data that shows a disk building period through 2000 to 2005, before dissipation in 2005, a seemingly quite variable period from 2005 to 2009, and then a final disk building event in 2010, reaching a quasi steady-state configuration beginning in 2011 to the present year. This behaviour is also seen in the \Halpha EW data of Figure \ref{fig:V_EW_plot}, where we see the clear building during the first few years of the century, a rapid dissipation of EW in 2005 to 2006, and a slight steady increase in EW from 2010 through to 2017. We do not see as dramatic of changes in EW as in the V-band, since the V-band comes from the inner portion of the disk, where mass loss and re-accretion occur whereas the \Halpha is produced in a larger disk volume and is slower to react to these episodic events \citep{Carciofi2011}. It is unclear whether the large variations in V magnitude seen from 2005 to 2010 are due to an affect from the secondary star, or some interaction solely between the primary star and its disk.

These building and dissipation phases are also confirmed by the parameters $\rho_0$ and $n$ of our HDUST models of \del Sco. In both cases of our modelling results (Figures \ref{fig:one_sigma_model_results} and \ref{fig:two_sigma_model_results}), we see our best-fit models start with a larger $\rho_0$ and high $n$ in the early century. The larger value of $\rho_0$ indicates more material in the inner disk, which leads to a bright V magnitude, and the high $n$ value means the disk density decreases rapidly with increasing radius, which leads to less material in the \Halpha forming region of the disk and a relatively low EW. This situation is reversed in our models for 2005 to 2007 where we see a lower value of $\rho_0$, which indicates a dimmer V magnitude, and a lower $n$ value, which leads to a higher ratio of material in the outer disk to the inner disk than in the case of a high $n$, and hence a larger \Halpha EW. Our best-fit models then show a transition period from 2009 to 2010, and come to a fairly constant combination of $\rho_0$ and $n$ for our disk models from 2011 to 2018. This is in perfect agreement with the theoretical expectation that the brightening of pole-on Be stars are the result of disk formation (high $\rho_0$), and the dimmings are associated with a (partial) disk dissipation (low $\rho_0$, see \citealt{Haubois2012}). The value of $n$ also seems to be in agreement with the picture outlined by \cite{Haubois2012} and \cite{Vieira2017}: larger values are usually associated with either brightening phases or phases of constant brightness, whereas low values of $n$ are seen in the dissipation phases. However, we acknowledge that for late-type Be stars the situation may be different.

The best-fitting inclination determined from our models is not constant. We see large variations of \ang{15} in our best-fitting models, and an even larger range of inclinations when including our errors. \cite{Cyr2017} showed that the disk of a Be star with a circular binary companion with an orbit misaligned by \ang{30} can cause a vertical tilting of the circumstellar disk by \ang{10}. If the companion of \del Sco has an orbit misaligned with the disk it is possible that this difference in inclination could be due to the binary companion, with the companion affecting the disk near periastron when it is close to the disk, and then the disk settling down when it is further from the primary star and disk, when the gravitational torque would not be as strong. The observational findings of \cite{tycner2011revised}, who found an orbital inclination of \ang{32.9} for \del Sco, indicate, when combined with our inclination results, that the disk and companion orbit are nearly coplanar. This does not support the hypothesis of possible disk tilting due to the companion star. However, it is entirely possible that the inclination of \del Sco cannot be well constrained with our methods, as there may be many degenerate combinations of $\rho_0$, $n$, and inclination that can produce the same simulated values of V magnitude and \Halpha EW.

Our best-fit HDUST models in Figure \ref{fig:two_sigma_model_plots} also display the variability of \del Sco since 2000. In each year of Figure \ref{fig:two_sigma_model_plots}, our models match the peak height of the average observed spectra reasonably well, however the overall shape of the model \Halpha spectral line does not always match perfectly. Looking at the spectra in Figure \ref{fig:two_sigma_model_plots} from 2009 to 2011, we see the red peak dominates the violet peak, while from 2014 to 2016, the violet peak dominates the red peak. This may indicate that the companion influences the disk and tidally locks a density enhancement in the disk as it passes close by the primary star, rotating from one side of the disk to the other. Due to this, the V/R variations of \del Sco require more detailed modelling efforts, which is beyond the scope and primary focus of this work. However, our symmetric models obtained here do capture the overall large scale variations of the disk of \del Sco.

Our modelling results also agree with what other modelling efforts have found. \cite{carciofi2006properties} found a $\rho_0$ of $4.5\times10^{-10}\, \rm g\, cm^{-3}$ in fitting their 2001 to 2004 photometry, which is of the same order of magnitude as our density of $1\times 10^{-10}\, \rm g\, cm^{-3}$ in 2000 to 2002. Their inclination of \ang{38} $\pm$ \ang{5} also is within the error of our best-fitting models for that time period. \cite{Arcos2017} fit an \Halpha spectra of \del Sco from 2014 and found a best-fitting model of  $\rho_0\,=\,7.5\times10^{-11}\, \rm g\, cm^{-3}$, $n\,=\,3$, and $i\,=\,$\ang{20}. Our best-fitting model for 2014 consisting of $\rho_0\,=\,1\times10^{-10}\, \rm g\, cm^{-3}$, $n\,=\,3.3$, and $i\,=\,$\ang{29} is convincingly close to their result, and the difference may be attributed to degeneracy in the different sets of parameters producing very similar simulated observations.

The radial extent of the \Halpha emitting region of \del Sco has been a subject of much focus in past publications. \cite{miroshnichenko2003spectroscopy} gave the first estimate for the size of the \Halpha emitting region as $10.8\, R_*$ in 2003, which is in agreement with the radius of our best-fitting models of 8 $ R_*$ in 2000 and 2001, as well as 14 $ R_*$ in 2002. They also state a mean outward expansion speed of 0.4 km $\rm s^{-1}$, or 2.67 $ R_*$/yr, which our radii numbers roughly agree with. \cite{millan2010spectro} determined the radius of their \Halpha emitting region to be 14.9 $\rm R_*$ through their 2007 observations, which is in agreement with our best-fitting 2007 model, whose radii we found to be 11 $ R_*$. \cite{meilland2011binary} also found a radius of 9 $R_*$ from their observations from 2007 to 2010. As with fitting the V/R ratios of the \Halpha spectra, the radii from these models need more investigation, and would most certainly benefit from more in depth modelling, accounting for possible density enhancements that would give similar measurements of \Halpha EW and V magnitude, while giving the disk a non-axisymmetric density structure. This structure could be in the form of spiral density enhancements as shown in \cite{Cyr2017}, and the shape of the disk could become very abstract around periastron, should the companion be in a retrograde orbit as shown in \cite{panoglou2016discs}.

The overall results of our modelling show an evolving picture of \del Sco. From forming early in the century, to dissipating halfway through the secondary's orbit, and finally building again before the most recent periastron, \del Sco has been very active since exhibiting its first strong sign of a circumstellar disk in 2000. However, it is unclear whether this activity can be attributed to the close passing of the companion star every 11 years. It will be of great interest to see how the system evolves leading up to the periastron in 2022 and beyond. Should we notice a large change from its now seemingly steady configuration, we will be able to confirm that the large companion star is having an effect on the circumstellar disk  of \del Sco. 

\section*{Acknowledgements}
\label{sec:acknowledgements}
We would like to thank the anonymous referee for their very thorough and detailed comments that improved the paper. C.E.J. and M.W.S. acknowledge support through the National Science and Engineering Research Council of Canada. G.W.H. acknowledges support from NASA, NSF, Tennessee State University, and the State of Tennessee through its Centers of Excellence program. A.C.C acknowledges the support from CNPq (grant 307594/2015-7). This work has made use of the computing facilities of the Laboratory of Astroinformatics (IAG/USP, NAT/Unicsul), whose purchase was made possible by the Brazillian agency FAPESP (grant 2009/54006-4) and the INCT-A. Ritter Observatory Public Archive is supported by the National Science Foundation Program for Research and Education with Small Telescopes (PREST). Based on observations obtained at the Canada-France-Hawaii Telescope (CFHT), operated by the National Research Council of Canada, the Institut National des Sciences de l'Univers of the Centre National de la Recherche Scientifique of France, and the University of Hawaii. We acknowledge with thanks the variable star observations from the AAVSO International Database contributed by observers worldwide and used in this research. This research has made use of the Vizier catalogue access tool, CDS, Strasbourg, France (DOI: 10.26093/cds/vizier). The original description of the Vizier service was published in A\&AS 143, 23. Based on INES data from the IUE satellite.

\software{HDUST \citep{carciofi2006non}, BEMCEE \citep{Mota2019}}

\bibliography{deltascobib}
\end{document}